\documentclass[11pt]{article}
\usepackage{amsmath,amsthm,amssymb}
\usepackage{graphicx}
\usepackage{tikz}
\usepackage{float}
\usepackage{hyperref}
\hypersetup{colorlinks,citecolor=blue,urlcolor=blue,linkcolor=blue}
\usepackage[citestyle=authoryear, backend=bibtex,natbib=true,firstinits=true,doi=false,isbn=false,url=false,eprint=false,maxbibnames=10]{biblatex}
\usepackage{bbm}
\usepackage{algorithm}
\usepackage{algorithmic}
\usepackage{caption}
\usepackage{subcaption}
\usepackage{hhline}
\usepackage{booktabs}
\usepackage{multirow}
\usepackage{array}
\usepackage{placeins}
\usepackage{makecell}

\usepackage{my_style}

\bibliography{ref.bib}

\newcommand{\ext}{\operatorname{ext}}
\newcommand{\pool}{\operatorname{pool}}

\newcommand{\FDR}{\operatorname{FDR}}
\newcommand{\sign}{\operatorname{sign}}
\newcommand{\bX}{\mathbf{X}}
\newcommand{\bY}{\mathbf{Y}}

\newcommand{\lc}{\operatorname{lro}}

\newcommand{\wl}{\operatorname{wl}}
\newcommand{\pval}{p\operatorname{-value}}
\newcommand{\pvals}{p\operatorname{-values}}

\makeatletter
\def\@fnsymbol#1{\ensuremath{\ifcase#1\or *\or 1 \or 2 \or 3 \or 4 \or 5 \else\@ctrerr\fi}}
\makeatother

\title{Transfer learning in genome-wide association studies with knockoffs}
\author{Shuangning Li\footnote{Authors listed alphabetically.} \thanks{Department of Statistics, Stanford University, Stanford CA 94305}, Zhimei Ren\thanks{Department of Statistics, University of Chicago, Chicago IL 60637}, Chiara Sabatti\thanks{Departments of Statistics and of Biomedical Data Science, Stanford University, Stanford CA 94305}, Matteo Sesia\thanks{Department of Data Sciences and Operations, University of Southern California, Los Angeles CA 90089}}

\begin{document}

\maketitle

\begin{abstract}
  This paper presents and compares alternative transfer learning methods that can increase the power of conditional testing via knockoffs by leveraging prior information in external data sets collected from different populations or measuring related outcomes. The relevance of this methodology is explored in particular within the context of genome-wide association studies, where it can be helpful to address the pressing need for principled ways to suitably account for, and efficiently learn from the genetic variation associated to diverse ancestries. Finally, we apply these methods to analyze several phenotypes in the UK Biobank data set, demonstrating that transfer learning helps knockoffs discover more numerous associations in the data collected from minority populations, potentially opening the way to the development of more accurate polygenic risk scores.
\end{abstract}

\section{Introduction}

Consider a supervised data set containing $p$ explanatory variables for each of $n$ individuals, $X \in \mathbb{R}^p$, and a corresponding outcome, $Y$, which may be either discrete or continuous. These data may be high-dimensional, in the sense that $p$ may be larger than $n$.
An interesting problem is to discover which variables in $X$ provide us with non redundant information on the value of $Y$: this knowledge represents a first step in understanding possible causal models linking $X$ and $Y$ and provides  the building blocks for robust prediction rules. 
This variable selection goal has been recently expressed in terms of multiple testing, by focusing on the 
 following null hypotheses of conditional independence \citep{candes2018panning} for all $j \in \{1,\ldots,p\}$:
\begin{equation}
\label{eqn:hypo_e}
H_j: X_j \indep Y \mid X_{-j},
\end{equation}
where $X_{-j}$ indicates all variables in $X$ except for the $j$-th one.

The method of {\em knockoffs}, introduced by \citet{barber2015controlling} for low-dimensional linear regression and later extended by \citet{candes2018panning} to the more general setting considered in this paper, allows one to test the hypotheses in~\eqref{eqn:hypo_e} and provably controls the false discovery rate (FDR)---the expected proportion of rejected nulls that are true \citep{benjamini1995}.
A distinctive advantage of conditional testing with knockoffs is that it is fully non-parametric, requiring no modeling assumptions about the distribution of $Y \mid X$, namely $P_{Y \mid X}$, which is possibly complex and generally unknown.
Instead, knockoffs treat the explanatory variables as random and model their joint distribution, $P_X$.
This approach is quite natural in some applications such as in a genome-wide association study (GWAS), where the goal is to discover which genetic variants, among the hundreds of thousands measured in modern studies, are useful to explain the inheritance of a polygenic trait (e.g., blood pressure) or disease (e.g., diabetes). 
In this context, knockoffs leverage well-established models for the transmission of genotypes from parents to offspring \citep{sesia2019gene} that provide a distribution $P_X$, without having to make assumption on the relation between genotypes and phenotypes, which  is typically unknown. 

While conditional testing with knockoffs leads to interpretable and meaningful discoveries in GWAS and other applications, the power of this methodology is naturally limited by the size of the sample at hand. In many context, however, scientists have available related data sets that can be leveraged to increase power. In genetics, for example, studies involving individuals of African descent are still limited in number and size \citep{popejoy2016}, but large collection of genotypes and phenotypes for subject of European ancestry are available \citep{bycroft2018} and can provide relevant information. 
Building upon prior work on knockoffs, this paper presents and compares different {\em transfer learning} methods for testing the conditional hypotheses $H_j$ in~\eqref{eqn:hypo_e} as powerfully as possible while leveraging prior knowledge from other data sets, which may be relevant but do not correspond exactly to the population of interest.
Although these methods will be generally applicable, this paper focuses in particular on their implications to the analysis of GWAS data,  where transferability across populations with different ancestries is of particular scientific and societal concern \citep{Duncan2019}, as we shall discuss after stating our statistical challenge more formally.

\subsection{Problem statement} 
% \color{red}
% Chiara: below I attempted to make the problem statement here a bit more general. I am likely going to make mistakes so you will need to look at this carefully and edit appropriately.
% To be honest, I do not know what is the appropriate choice. While we are talking about different environments, there clearly should be something that links these models together for one to even want to start doing transfer learning. On the other hand, if we have it as is, where $Y$ and $X$ are the same across environments, we do not quite account for the case where we look at different phenotypes in different populations. I ended up only using only words, but I think it would be important to pay attention at what we index with $e$: just the $P_{Y|X}$ or the $Y$ and $X$ themselves.

% I also removed the part about filtering at the end as I think it is not needed and it might be difficult for people to understand said so quickly

% \color{black}

Imagine that data were collected from many {\em environments} (specific studies, experimental settings, or distinct sub-populations) $e \in \{0,1,\ldots,E\}$, for some $E \in \mathbb{N}$, each corresponding to a particular distribution $P^e_{XY} = P^e_{X} \cdot P^e_{Y \mid X}$, where $P^e_{X}$ denotes the joint distribution of the explanatory variables and $P^e_{Y \mid X}$ is the conditional distribution of the outcome. The object of inference is $P^e_{Y \mid X}$ in the {\em target} environment $e=0$. That is, we wish to test the conditional hypotheses $H_{j}$ in~\eqref{eqn:hypo_e} corresponding to the unknown $P^e_{Y \mid X}$ for $e=0$; we will refer to this specific null as~$H_{j}^0$. 
Even though both $P^e_{Y \mid X}$ and the definition of the $X,Y$ variables may differ across environments, we are interested in situations where one has reason to believe that there might be commonalities across the $P^e_{Y \mid X}$, so that it makes sense to attempt learning from others. 

It is useful to focus on a specific instance of the problem above and contrast the goal of our analysis with those of other existing approaches.  
Let us begin by assuming the definitions of the $X$ and $Y$ variables are the same in all environments, so that there is a clear correspondence between the various $H_{j}^e$ for all $e$. Our goal is to leverage all available information (transfer learning) to test $H_{j}^0$: this makes sense when one believes that the environment $e=0$ is sufficiently distinct from the others and of specific scientific interest. By contrast, in other contexts one may want instead to test the specific hypotheses $H_{j}^e$ separately environment-by-environment using the original method of \citet{candes2018panning}, or the
global null $\mathcal{H}^{\text{meta}}_{j} :\cap_{j=0}^E H_{j}^e$, as it happens in meta-analysis studies, or the composites null
%\begin{align} \label{eq:null-ici}
$\mathcal{H}^{\text{comp}}_{j} : \exists e \in \{1,\ldots, E\} \text{ such that the null } \mathcal{H}^{e}_{j} \text{   is true} $,
%\end{align}
such as in \citet{li2021searching}, or softer partial conjunction \citep{benjamini2008screening} versions of the latter.

The transfer learning problem considered in this paper is non-trivial; for example, the standard knockoffs methodology of \citet{candes2018panning} applied to the pooled data from all environments would not be a fully satisfactory solution to test $H_{j}^0$,  although it is a valid and reasonable test of $\mathcal{H}^{\text{meta}}_{j}$.
Indeed, $P^e_{Y \mid X}$ may vary across environments, and in that case the above naive approach would be invalid because we seek inferences about $\smash{ P^0_{Y \mid X}} $, not on some mixture distribution. 
Further, even if  $\smash{P^{e}_{Y \mid X}}$ were to identify the same non-null variable for all $e \in \{0,\ldots,E\}$, in which case pooling would provide a valid test of $H_j^{0}$ within the target environment, there may be at least two compelling motivations to search for alternative methods of analysis. 
First, one may have direct access only to the observations from the target environment, with the information from the other data sets compressed in the form of summary statistics, due to privacy or computing constraints, for example.
Second, {\em covariate shifts} (changes in $\smash{P^{e}_X}$) may render some discoveries more or less relevant in different environments. Concretely, imagine a variable $X_j$ that in theory has the same effect on $Y$ according to all $\smash{ P^{e}_{Y \mid X}} $, but in practice is always (nearly) constant for all observations from the target environment. If $X_j$ varies appreciably in the other environments (which have different $\smash{P^e_{X}}$), we should expect to discover it through pooling. However, this finding would not be very useful if our focus is to predict $Y$ from $X$ for future observations from the target environment.

\subsection{Transferability in GWAS}

While the transfer learning methods studied in this paper are widely applicable, our work is particularly motivated by a specific problem in GWAS.
The problem arises from the fact that most of the genotype and phenotype data collected so far and available for general analysis involve individuals of European descent, either living in Europe or in the United States. 
This introduces some biases and represents a challenge to the development of clinically useful prediction tools based on genetic information \citep{popejoy2016,sirugo2019missing}. It is therefore important to analyze the relatively scarce data available from diverse populations in the most efficient way, possibly leveraging, or ``transferring'', some of the information already gathered from Europeans in order to discover the most relevant genetic variants for the minority populations. To avoid misunderstanding, and to clarify what assumptions might be appropriate in this context, it is useful to provide some additional background.

As a consequence of the processes by which human populations evolved, many genetic variants have different allele frequencies in groups with different ancestries, and the dependency patterns within one chromosome (linkage disequilibrium) also vary across populations \citep{laan1997demographic}.
For example, a variant might have originated five generations ago in an individual living in Finland: in this case we would expect to see the alternative allele only in this person's descendants, or at least predominantly so as new mutations are always possible. 
Or it might be that an allele was common in the handful of individuals who left Africa to populate a region in Asia: this allele will now have higher frequency among individuals who originate from that region, and it will be associated with alleles at neighboring variants that were present in the founder groups. 
By contrast, the same allele might have a different set of associated neighboring alleles in populations descendant from other lineages.

Differences in allele frequencies and in linkage disequilibrium patterns are such that the genotypes of each individual carry distinctive signatures of the human populations to which they belong \citep{rosenberg2002}. 
This must be kept in mind by geneticists in order to avoid making an excessive number of false positives while analyzing GWAS data \citep{Price2006}, and it has in part motivated the use of ethnically homogeneous samples. 
However, as the field of statistical genetics progresses and more associated variants are discovered, the bias towards the analysis of data from individuals of European descent gives rise to more issues. In particular, it has been observed that the genetic markers with discovered associations tend to have better predictive power and association strength in Europeans \citep{wray2013,martin2017,Duncan2019} compared to other populations, to the point that it has been suggested that their use in clinical practice might be unequitable \citep{martin2019}. 
In response to this problem, there have several recent efforts to collect GWAS data from more diverse populations \citep{wojcik2019}, and to develop new statistical methodologies that incorporate them effectively \citep{coram2017}.

The importance of purposefully collecting data from individuals that identify as members of different ethnicities \citep{popejoy2016,Reich2018}, as well as of the appropriateness of developing population-specific models for disease risk prediction, have lead to some misinterpretations and controversy \citep{holmes2018}. 
Note that the term ``population'' is used here loosely to indicate a group of people with common ancestry that mates prevalently internally and shares a common environment and lifestyle. Of course, populations are becoming much more intermixed in our modern society, but some of the meaningful variations associated with their historical roots persist.
The medical relevance of genetic variation can be understood by imagining that the same biological processes are responsible for translating genotypes to phenotypes for all humans, irrespective of ancestry. 
However, while attempting to reconstruct through statistical analyses this underlying causal model, and to understand the roles played within it by different genetic variants, we are faced with particular difficulties due to the variability in allele frequency, linkage disequilibrium, and environmental exposures that characterizes human populations. 
For example, a causal variant might occur in a particular population rarely, or even not at all. 
Further, a causal variant might affect the phenotype by interacting with other non-genetic variables, i.e., consumption of gluten. 
The level of gluten intake in different populations can be quite diverse, and so one should expect to observe different effect sizes for the aforementioned genetic variant in models of the phenotype that do not account for gluten consumption.
Finally, a causal variant may not be directly measured in the data set at hand, and the genotyped proxies which best capture its underlying signal may differ across population depending on their patterns of linkage disequilibrium.

The above considerations suggest that one should carefully take into account the ancestry of each individual while analyzing GWAS data from heterogeneous populations. 
Indeed, the importance of population-specific approaches has been underscored in a number of papers demonstrating that models fitted on data from European samples do not predict accurately the phenotypes of individuals with other ethnicities \citep{wray2013,martin2017,Duncan2019}. 
At the same time, given the scarcity of data on individuals of non-European descent, and based on the principle that the fundamental underlying biology is the same for all humans, it is also important to use models that can leverage the information available in European samples when analyzing data from other populations, and this is precisely the goal of ``transfer learning.''

\subsection{Related work}

While Bayesian inference presents a natural framework to leverage side information, we are here interested in working within the frequentist framework, and specifically through hypothesis testing. Previous work has illustrated how to incorporate 
side information within the context of hypothesis testing with FDR control, mostly focusing  in settings in which p-values are available. For example, \citet{genovese2006false} developed a weighted variation of the Benjamini-Hochberg (BH) procedure \citep{benjamini1995} such that larger prior weights make it easier for the corresponding hypotheses to be rejected. 
Later works presented different ways of determining the weights, either assuming the side information is independent of the $\pvals$ \citep{roquain2009optimal}, or allowing for data dependent weighting \citep{hu2010false, zhao2014weighted, ignatiadis2016data, durand2019adaptive}. The weighted BH procedure was further extended by \citet{ignatiadis2017covariate}, which considered learning data-adaptive $\pval$ weights using cross-weighting, and by \citet{lei2018adapt}, which trained a sequence of $\pval$ thresholds adaptively and iteratively. 

If valid $\pvals$ are not available, as it is the case for conditional testing with GWAS data \citep{sesia2018,sesia2020}, the above methods cannot be applied.
To address this challenge, \citet{ren2020knockoffs} developed an adaptive knockoff filter, extending the original knockoff filter of \citet{barber2015controlling} and \citet{candes2018panning} in order to increase its power by leveraging side information. This paper applies the method of \citet{ren2020knockoffs} to analyze GWAS data from individuals with different ancestries \citep{sesia2020controlling}, comparing its performance to that of a novel alternative approach. The main difference between the adaptive knockoff filter of \citet{ren2020knockoffs} and the novel method proposed here in Section~\ref{subsection:adaptive} is that the latter directly leverages side information while analyzing the raw genotype-phenotype data, while the former operates on pre-computed knockoff statistics.
Our solution is made possible by the general flexibility of the knockoffs framework \citep{candes2018panning}, which allows one to control the FDR utilizing any test statistics. This work is also partially inspired by \citet{li2021searching}, which study the related problem of testing for robust associations that are consistent across many environments.

The idea of leveraging side information from external data sets is common in the context of predictive inference, where it is typically described as ``transfer learning'' \citep{heckman1979sample,pan2009survey}. %Transfer learning is particularly challenging in genetics, and it is well known that polygenic risk scores do not always perform well across populations with different ancestries \citep{Duncan2019}. 
While this paper focuses on testing rather than prediction, these problems are closely related and the methods described here could naturally be applied to construct predictive models, for example selecting which genetic markers should be utilized to compute more efficient polygenic risk scores leveraging the information contained in GWAS data from different populations.

\section{Methods} \label{section:methods}

\subsection{Review: knockoffs}
\label{subsection:knockoffs_review}

Knockoffs provide a method to analyze data from one environment $e$ and test the conditional hypothesis $H_j^e$~\eqref{eqn:hypo_e}, controlling the FDR over all variables $j \in \{1,\ldots,p\}$.
In the model-X setting we consider \citep{candes2018panning}, the first step of this procedure consists of generating synthetic variables $\tilde{X}^e = (\tilde{X}_1^e, \dots, \tilde{X}_p^e)$ that imitate the distribution of the original variables $X^e = (X_1^e, \dots, X_p^e)$ but are known to be null in the sense of $H_j^e$~\eqref{eqn:hypo_e}. In particular, $\tilde{X}^e$ is created as a function of $X^e$ without looking at $Y^e$, so that $\tilde{X}^e \indep Y^e \mid X^e$, and it is carefully designed to ensure that the joint distribution of $(X^e,\tilde{X}^e)$ remains unaltered if any variables are swapped with the corresponding knockoffs: $\smash{(X^e, \tilde{X}^e)_{\operatorname{swap}(S)} \stackrel{d}{=}(X^e, \tilde{X}^e)}$ for any $\mathcal{S} \subset \cb{1, \dots, p}$.
Here, the notation $\smash{(X^e, \tilde{X}^e)_{\operatorname{swap}(S)}}$  indicates the $2p$-dimensional vector obtained by swapping the elements of $X^e$ indexed by $S$ with the corresponding elements of $\tilde{X}^e$. In practice, the construction of knockoffs requires knowledge of the joint distribution of the original variables, $P_X^e$, and practical algorithms have already been developed to handle many possible cases, including multivariate Gaussian distributions \citep{candes2018panning} and hidden Markov models \citep{sesia2019gene}. Even if $P_X^e$ is completely unknown, algorithms are available to construct approximate knockoffs \citep{romano2020deep}.
As our work focuses on a separate aspect of the analysis, we assume henceforth that valid knockoffs are available.
 
The second step of a knockoff analysis computes test statistics $W^e \in \mathbb{R}^p$ that provide information on the potential dependency of $Y$ on $X$. These statistics are a function of all data $\bX^e \in \mathbb{R}^{n \times p}, \bY^e \in \mathbb{R}^n$ and knockoffs $\tilde{\bX}^e \in \mathbb{R}^{n \times p}$, and they encode the idea that the original variables must be significantly more predictive of $Y$ compared to $\tilde{X}$ in order to allow the rejection of the null $H_j^e$~\eqref{eqn:hypo_e}.
Specifically, each element of $W$ is defined as
\begin{equation}
W^e_{j}=w_{j}([\bX^e, \tilde{\bX}^e], \bY^e),
\end{equation}
where the function $w_j$ satisfies the following flip-sign property: swapping the $j$th column of $\bX^e$, namely $\bX_j^e$, with $\tilde{\bX}_j^e$ has the only effect of changing the sign of $W_j^e$. 
A typical way of computing these statistics is to estimate a sparse linear (or generalized linear) regression model of $\bY^e$ given the (standardized) predictors $[\bX^e, \tilde{\bX}^e]$, extract the fitted coefficients $\hat{b}^e \in \mathbb{R}^{2p}$, and set $W^e_j = |\hat{b}_j^e| - |\hat{b}^e_{j+p}|$.
If $H_j^e$~\eqref{eqn:hypo_e} is not true, one expects to see a large $\smash{|\hat{b}_j^e|}$ but small $\smash{|\hat{b}_{j+p}^e|}$, because $\smash{\tilde{\bX}^e_j}$ is by construction independent of $\bY^e$ given the other variables; in that case, the corresponding $W^e_j$ would tend to be positive and large.
By contrast, if $H_j^e$~\eqref{eqn:hypo_e} is true, $[\bX^e_j, \tilde{\bX}^e_j]$ has the same distribution as $[ \tilde{\bX}^e_j, \bX^e_j]$ conditional on $\bY^e$ and on the other variables, and thus $W^e_j$ is equally likely to be positive or negative.
Formally, the signs of the null $W^e_j$ are i.i.d.~coin flips conditional on $(|W^e_1|, \dots, |W^e_p|)$ \citep{candes2018panning}, and this allows the computation of a significance threshold guaranteeing FDR control below any desired level $q \in (0,1)$.
This threshold is computed by the knockoff filter \citep{barber2015controlling} as
\begin{equation} \label{eq:knockoff-filter}
T^e=\min \left\{t: \frac{1+\#\left\{j: W^e_{j} \leq-t\right\}}{\#\left\{j: W^e_{j} \geq t\right\} \vee 1} \leq q\right\},
\end{equation}
and the corresponding set of discoveries is $\hat{\mathcal{S}} = \{j: W^e_j \geq T^e\}$, with $\min \emptyset = \infty$.

An equivalent reformulation of the knockoff filter~\eqref{eq:knockoff-filter} which will be useful in this paper is the following. Imagine sorting the $p$ target hypotheses in ascending ordering of the absolute values of the knockoff statistics $W^{e}$ input to the filter, i.e., $|W^{e}_{\pi_{1}}| \leq |W^{e}_{\pi_2}| \leq \ldots \leq |W^{e}_{\pi_{p}}|$, where $\pi_1, \ldots, \pi_p$ indicate the order statistics of $|W^{e}|$. Then, sequentially compute 
\begin{equation} \label{eq:knockoff-filter-2}
\widehat{\FDR}(k) = \frac{1+\sum_{j > k} \mathbf{1} {\left\{W^{e}_{\pi_{j}}<0\right\}}}{\left(\sum_{j > k} \mathbf{1} {\left\{W^{e}_{\pi_{j}}>0\right\}}\right) \vee 1},
\end{equation}
for each $k = 0,1, \dots, p-1$ until $\widehat{\FDR}(k) \leq q$, and reject all $H_{\pi_j}$  such that $j >  k$ and  $W^{e}_{\pi_{j}}>0$. In the case that $\smash{\widehat{\FDR}(k) > q}$ for all $k$, no hypotheses are rejected.
This formulation highlights that the role of $|W^{e}_j|$ is to offer an informative ordering of the hypotheses, with the idea that those with positive statistics should be found at the end of this sequence in order to maximize power.

\subsection{Transfer learning with linearly re-ordered knockoff statistics}
\label{subsection:linear_com_stats}

If the goal is to use knockoffs to test $H_j^e$~\eqref{eqn:hypo_e} for one particular environment, i.e., $e = 0$, the trivial solution is to apply the existing methodology reviewed in the previous section to the available data collected from the population of interest \citep{candes2018panning}.
However, if additional observations are available from other environments $e \in \{1,\ldots, E\}$ which may share some similarity in $P_{Y \mid X}$, one may want to incorporate that information into the analysis as efficiently as possible.
To this end, consider the following simple procedure: after generating knockoffs for every $e \in \{0, 1,\ldots,E\}$, apply the standard method from \citet{candes2018panning} to the data from environment $e=0$ obtaining statistics $W^0$, and to the pooled data from the external environments $e \in \{1,\ldots,E\}$ obtaining statistics $W^{\ext}$.
Then, combine $W^0$ and $W^{\ext}$ into new statistics $W^{\lc} \in \mathbb{R}^p $ defined such that:
\begin{equation}
\label{eqn:lc_sign} 
\sign (W_{j}^{\lc})= \sign (W_{j}^{0}).
\end{equation}
\begin{equation}
\label{eqn:lc_magnitude} 
|W_{j}^{\lc}|= (1-\theta) |W_{j}^{0}| +  \theta |W_{j}^{\ext}|, \quad \text{ for some fixed } \theta \in [0,1]. 
\end{equation}
In words, the signs of $W^{\lc}$ are the same as those of the naive statistics computed on the environment of interest, while their absolute values are a linear combination of $|W^{0}|$ and $\smash{|W^{\ext}|}$. In the special case of $\theta = 0$, we recover $W^{\lc} = W^{0}$.
This procedure ensures the signs of null $W^{\lc}$ are i.i.d.~coin flips conditional on $|W^{\lc}|$, which implies the standard knockoff filter~\eqref{eq:knockoff-filter} can be applied to control the FDR for the hypotheses $H_j^e$~\eqref{eqn:hypo_e} in the environment $e=0$ of interest \citep{candes2018panning}.
\begin{prop}
\label{prop:coin_lc}
Let $W^{\lc}$ be the knockoff statistics for the target environment $e = 0$ computed with the linear re-ordering method described above, based on prior importance statistics $W^{\ext}$ computed on data from the external environments $e \in \{1,\ldots,E\}$.
Conditional on $|W^{\lc}|$, the signs of $W_j^{\lc}$ for all null $j$ corresponding to a true $H_{j}^0$~\eqref{eqn:hypo_e} are i.i.d.~coin flips.
\end{prop}

Intuitively, if the null $H_j^e$~\eqref{eqn:hypo_e} is not true for $e=0$ and the external environments are similar to the target one, the value of $\smash{W^{\ext}_j}$ tends to be larger than that of $\smash{W^{0}_j}$. This increases power, especially if the combined sample sizes for the other environments are larger than that of the target one.
By contrast, if $X_j$ is null in the sense of $H_j^e$~\eqref{eqn:hypo_e} for $e=0$, the sign of $W^{\lc}_j$ is still a fair coin flip independent of everything  else (the statistics computed on data from different environments are mutually independent), which allows rigorous FDR control.
The value of $\theta$ in \eqref{eqn:lc_magnitude} must be specified before looking at the data, but unfortunately the optimal choice that maximizes power depends on the data. For example, one should intuitively choose a larger $\theta$ if the external environments have large sample sizes and are similar to the target one, while smaller values of $\theta$ may otherwise be preferable. This dilemma motivates the following alternative approach.

\subsection{Transfer learning with the adaptive knockoff filter}
\label{subsection:adaptive}

The work of \citet{ren2020knockoffs} developed an extension of the knockoff filter that can be directly applied to address our transfer learning problem. The advantage of this {\em adaptive knockoff filter} is that it can leverage the external statistics $W_{j}^{\ext}$, and any other relevant prior information, in a more flexible and data-driven manner, possibly yielding higher power without involving the particularly sensitive and unknown parameter $\theta$ required by the latter.
In particular, \citet{ren2020knockoffs} prove that the procedure in~\eqref{eq:knockoff-filter-2} controls the FDR even if the statistics are re-arranged in some other data-dependent order $\pi$, as long $\pi$ satisfies a sign-invariance property that generally allows much more flexibility compared to the original knockoff filter~\eqref{eq:knockoff-filter}.
Informally, their approach dynamically learns at each step $k$ of~\eqref{eq:knockoff-filter-2} a new data-dependent ordering $\pi^k$, combining the prior information contained in $|W|$ and $W_{j}^{\ext}$ with the additional knowledge of the signs of $W_{\pi_j}$ for all $j \leq k$. This solution can adaptively adjust the weight given to the external information, relative to that given to the internal statistics, based on preliminary estimates of the numbers of discoveries that may be achievable on the available data set. As a result, if the side information is relevant, their procedure tends to relocate more of the positive statistics at the end of the testing sequence, thereby increasing the number of rejections.
%We refer to this method as the {\em adaptive knockoff filter} and refer to \citet{ren2020knockoffs} for further details.

While the adaptive knockoff filter is in principle very flexible, we highlight here a particular implementation that extends intuitively the linear combination approach outlined in \eqref{eqn:lc_sign}--\eqref{eqn:lc_magnitude}.
In fact, the aforementioned method from the previous section is equivalent to applying the procedure in~\eqref{eq:knockoff-filter-2} with $\pi_{k + 1} = \argmax_{j \notin \cb{\pi_1, \dots, \pi_k}} \{  (1-\theta) |W_{j}^{0}| +  \theta |W_{j}^{\ext}| \}$, for a parameter $\theta$ fixed a priori. 
The adaptive knockoff filter generalizes this solution as it allows tuning a different value of $\theta$ at each step $k$. For example, consider the following logistic model for the unknown signs of the test statistics:
\[ \operatorname{logit} \mathbb{P}[ \sign(W_j^{0}) = -1 ] = \theta_1 |W_{j}^{0}| + \theta_2 |W_{j}^{\ext}|,\]
for some parameters $\theta_1, \theta_2$.
The adaptive knockoff filter fits this model on the data in $\smash{ \{ W^{\ext}, |W^{0}_j| \} }$ and $\smash{ \{W_j^{0}: j \leq k\} }$, yielding updated estimates $\hat{\theta}_1$ and $\hat{\theta}_2$ at each step $k$. Then, the $(k+1)$-th hypothesis tested by~\eqref{eq:knockoff-filter-2} is that deemed most likely to be negative; i.e.,  $\pi_{k + 1} = \argmax_{j \notin \cb{\pi_1, \dots, \pi_k}} \big(\hat{\theta}_1|W_{j}^{0}| +  \hat{\theta}_2 |W_{j}^{\ext}|\big)$. Consequently, the positive statistics tend to be pushed towards the end of the sequence, increasing power while maintaining FDR control. 
Of course, in general there is no particular reason for the adaptive knockoff filter to rely on this simple logistic model; instead, any model can be exploited at the $k$-th step to predict the remaining unknown statistics signs from the data in $\{ W^{\ext}, |W^{0}_j| \} \cup \{W_j^{0}: j \leq k\}$ and any other relevant prior knowledge.

% Adaptive knockoffs is a very flexible procedure; other than the linear combination statistics, there are many other algorithms one can use to train the next hypothesis $\pi_{k+1}$. 
% To put it in a form similar to \eqref{eqn:lc_sign} and \eqref{eqn:lc_magnitude}, we note that adaptively knockoffs can be understood as 
% \begin{equation}
% \label{eqn:adaptive_sign} 
% \sign \left(W_{j}^{\adp}\right)= \sign \left(W_{j}^{0}\right).
% \end{equation}
% \begin{equation}
% \label{eqn:adaptive_magnitude} 
% \abs{W^{\adp}}= f\p{ \sign (W^{0}), |W^{0}| , |W^{\ext}| },
% \end{equation}
% for some function $f$ whose corresponding ordering function satisfies the sign invariant property. 

\subsection{Transfer learning with prior-informed knockoff statistics}
\label{subsection:penalty}

The transfer learning approaches described in Sections~\ref{subsection:linear_com_stats}--\ref{subsection:adaptive} are based on statistics whose signs are determined entirely by the observations in the target environment, independently of all external data. Indeed, the only difference between these two methods is the order in which they filter the statistics~\eqref{eq:knockoff-filter-2}. However, the flexibility of the knockoffs framework allows the external data to also inform the signs of the test statistics, possibly further increasing power. In fact, it is well-known that any available prior information can be directly incorporated into the predictive model of $\bY \mid \bX, \tilde{\bX}$ utilized to compute the test statistics input to the standard knockoff filter \citep{candes2018panning}, since any model can be employed for this purpose. However, it is unclear how to best take advantage of such extreme flexibility in order to maximize power.
Below, we present a concrete implementation of this procedure which intuitively generalizes the sparse generalized linear model (lasso) statistics reviewed in Section~\ref{subsection:knockoffs_review} and often performs well in practice; this approach takes inspiration from the recent work of \citet{li2021searching} which studied the related problem of testing for associations that are consistent across many environments.

Consider fitting a sparse generalized linear regression model of $\bY^0 \mid \bX^0, \tilde{\bX}^0$ with feature-specific $\ell_1$ regularization parameters $\lambda_j > 0$ defined for $j \in \{1,\ldots,2p\}$ as:
\[
  \lambda_j = (1-\gamma) \lambda + \gamma \phi_j.
\]
Above, $\lambda>0$ and $\gamma \in (0,1)$ are hyper-parameters tuned by cross-validation, and $\phi_j>0$ is a symmetric inverse measure of the prior importance of $X_j$ or $\tilde{X}_j$ satisfying $\phi_j = \phi_{j + p}$ for all $j \in \{1,\ldots,p\}$. In particular, $\phi_j$ should take smaller values for more promising variables; more details about how this can be computed will be provided later. 
The estimated regression coefficients for the original variables and knockoffs are then combined pairwise as in Section~\ref{subsection:knockoffs_review} to obtain the final ``weighted-lasso'' knockoff statistics, to which we refer as $W_{j}^{\wl}$.
In the special case of $\gamma = 0$, or $\phi_j = 1$ for all $j$, this reduces to the standard solution that does not leverage any external information. 
As $\gamma$ is tuned to maximize the predictive accuracy of the model within the target environment, this parameter should be close to zero if the external information is not helpful; by contrast, larger values $\gamma$ will tend to be selected if the data from the other environments are relevant. 
In the latter case, we expect truly important variables to receive weaker regularization and thus be more likely to contribute actively to the sparse model, thereby allowing us to reduce the noise level, possibly resulting in more powerful statistics. The symmetry requirement that $\phi_j = \phi_{j + p}$ for all $j \in \{1,\ldots,p\}$ ensures the final statistics satisfy the usual flip-sign property \citep{candes2018panning} necessary to guarantee FDR control with the knockoff filter.

\begin{prop}
\label{prop:coin_wl}
Let $W^{\wl}$ be the knockoff statistics for the target environment $e = 0$ computed with the weighted lasso method described above, based on prior importance weights $\phi$ computed on data from the external environments $e \in \{1,\ldots,E\}$.
Conditional on $|W^{\wl}|$, the signs of $W_j^{\wl}$ for all null $j$ corresponding to a true $H_{j}^0$~\eqref{eqn:hypo_e} are i.i.d.~coin flips.
\end{prop}

The above FDR guarantee allows the prior weights $\phi_j$ to be completely arbitrary, as long as they only depend on the external data from other environments or on other prior knowledge. As a concrete example, we consider $\smash{ \phi_j = 1 / (0.05 + |\hat{b}^{\ext}_j| + |\hat{b}^{\ext}_{j+p}| ) }$ for $j \in \{1,\ldots,p\}$, where $\smash{ \hat{b}^{\ext}_j }$ and $\smash{ \hat{b}^{\ext}_{j+p}} $ are the estimated coefficients of the (scaled) variables $X_j$ and $\tilde{X}_j$, respectively, in a sparse regression model fitted on the external data. Of course, there is no true need here to analyze the knockoffs from the other environments because the original variables already contain all possible relevant knowledge. Therefore, an equally valid simpler alternative would be to compute the above $\phi_j$ without including $\smash{ |\hat{b}^{\ext}_{j+p}| }$. 
Nonetheless, we will adhere to the current choice in this paper because it is convenient if a previous knockoff analysis was carried out on the data from the external environments, in which case one can simply recycle the already fitted regression model, and it is particularly useful to introduce an interesting variation of this method discussed below, in which the explicit inclusion of the knockoffs becomes important.

% Recall that a typical way of computing feature statistics $W_j$ is through lasso. 
% To obtain more informative feature statistics $W_j$ from side information, instead of directly working with statistics $W^e$ and $W^{\ext}$ as in the previous sections, an alternative way is to use $W^{\ext}$ in the process of fitting lasso. We compute feature statistics $W^{\ext}$ as above. Then for the target population $e$, we generate knockoffs $\widetilde{X}^e$ and run lasso with a specific penalty factor $\pi_j$:
% 	\[\hat{\beta} = \argmin_{\beta} \sum_i\p{y_i^e - \p{X_i^e,\widetilde{X}^e_i}^T \beta}^2 + \lambda \sum_{j = 1}^p \p{\abs{\beta_j} + \abs{\beta_{j+p}}} \pi_j, \]
% where we take 
% \[\pi_j = \p{1 - \gamma} + \gamma \frac{1}{0.05 + \abs{W_j^{\ext}}}. \]
% Here we tune the value of $\gamma$ and $\lambda$ by cross validation. 
% We then take the feature statistics to be
% \[W^{\wl}_j = \abs{\beta_j} - \abs{\beta_{j + p}}. \]

In the special case where the sets of nulls in the sense of $H_j^e$~\eqref{eqn:hypo_e} are the same for all environments $e \in \{0,\ldots,E\}$, even more flexibility is allowed in the construction of the prior weights $\phi$. 
For example, consider setting $\smash{ \phi_j = 1 / (0.05 + |\hat{b}^{\pool}_j| + |\hat{b}^{\pool}_{j+p}| ) }$, for $j \in \{1,\ldots,p\}$, where $\smash{ \hat{b}^{\pool}_j }$ and $\smash{ \hat{b}^{\pool}_{j+p}} $ are the estimated coefficients of the variables $X_j$ and $\tilde{X}_j$, respectively, in a sparse regression model fitted on the pooled data obtained by combining the external sets ($e\ \in \{1,\ldots,E\}$) including also the observations from the target environment ($e=0$). Then, imagine computing test statistics $W^{\wl}$ by applying the weighted lasso method described above to this $\phi$.
If the null variables are the same in all environments, the result of Proposition~\ref{prop:coin_wl} still holds even though $\phi$ is not independent of the data in the target environment, as established by the next proposition. The intuition behind this result is that the proof of Proposition~\ref{prop:coin_wl} can be suitably modified by randomly swapping null variables with their corresponding knockoffs simultaneously in all environments instead of acting only within the target one; however, this leaves the joint distribution of $[\bX,\tilde{\bX}, \bY]$ invariant, as required by the knockoff filter to ensure FDR control \citep{candes2018panning}, if and only if the variables which are null for $e=0$ are also null for all other $e \in \{1,\ldots,E\}$.

\begin{prop}
\label{prop:coin_wl-2}
Let $W^{\wl}$ be the knockoff statistics for the target environment $e = 0$ computed with the weighted lasso method described above, based on prior importance weights $\phi$ computed on the pooled data set obtained by combining the observations from all environments $e \in \{0, 1,\ldots,E\}$.
Assume all variables which are null in the target environment are also null in all other environments $e \in \{1,\ldots,E\}$.
Then, conditional on $|W^{\wl}|$, the signs of $W_j^{\wl}$ for all null $j$ corresponding to a true $H_{j}^0$~\eqref{eqn:hypo_e} are i.i.d.~coin flips.
\end{prop}

\section{Numerical Experiments} \label{sec:experiments}

A software implementation of our methods is available online at \url{https://github.com/lsn235711/transfer_knockoffs_code}, along with code to reproduce the analyses.

\subsection{Synthetic data}

We begin by comparing empirically the performances of the different approaches to transfer learning from Sections~\ref{subsection:linear_com_stats}--\ref{subsection:penalty} on synthetic data. Here, the adaptive knockoff filter is run with the ``gam" filter \citep{ren2020knockoffs}, while the weighted-lasso knockoff statistics are obtained with $\smash{ \phi_j = 1 / (0.05 + |\hat{b}^{\pool}_j| +  |\hat{b}^{\pool}_{j+p}| ) }$, for $j \in \{1,\ldots,p\}$, where $\smash{ \hat{b}^{\pool}_j }$ and $\smash{ \hat{b}^{\pool}_{j+p}} $ are defined as in Section \ref{subsection:penalty}.
As it is not clear a priori how to best tune the parameter $\theta$ needed by the linearly re-ordered knockoff statistics, we utilize an imaginary oracle to select the value of $\theta$ yielding the largest number of discoveries in each experiment. Of course, this is not guaranteed to control the FDR in theory and hence it may not be a valid approach in practice, but it provides an informative comparison with the other two methods within the scope of these simulations.
As benchmarks, we consider the following two approaches: (i) the {\em vanilla} knockoffs analysis \citep{candes2018panning} applied only to the data from target environment; and (ii) the {\em pooling} heuristic in which the standard knockoffs methodology is applied to the pooled data from all environments. Both benchmarks are applied using the standard lasso-based statistics reviewed in Section~\ref{subsection:knockoffs_review}, as in \citet{candes2018panning}. 

Simulated data are obtained from 3 different environments, consisting of $p = 500$ variables and $n = 800$ observations per environment.
In all environments, the variables are generated from an autoregressive model of order one with correlation parameter $\rho = 0.5$, and the knockoffs are constructed based on the true $P_X^e$ with the standard algorithm from \citet{candes2018panning}.
The conditional distribution of $Y^e \mid X^e$ in the $e$-th environment is given by $Y^e = X^e \beta^e + \epsilon^e$,
where $\beta^e \in \mathbb{R}^p$ is an environment-specific effect parameter vector, and $\epsilon^e$ are i.i.d.~standard Gaussian noise. Note that here the model generating the outcome $Y$ varies across environments.
In each environment, 60 entries of $\beta^e$ are equal to $a/\sqrt{n}$, with $a = 3.5$, while the others are zero.
The sets of non-zero entries of $\beta$ in each environment, $S^0, S^1, S^2 \subseteq \{1,\ldots,p\}$  respectively, are chosen at random such that $S^1 = S^2$, while the overlap (the proportion of shared elements) between $S^0$ and $S^1$ is varied as a control parameter.
Our goal is to find which variables are non-null in the first environment (i.e., to discover $S^0$), controlling the false discovery rate below $10\%$.
All experiments are repeated 500 times, averaging the empirical false discovery proportion and power.

\begin{figure}[!htb]
 \centering
 \includegraphics[width=\linewidth]{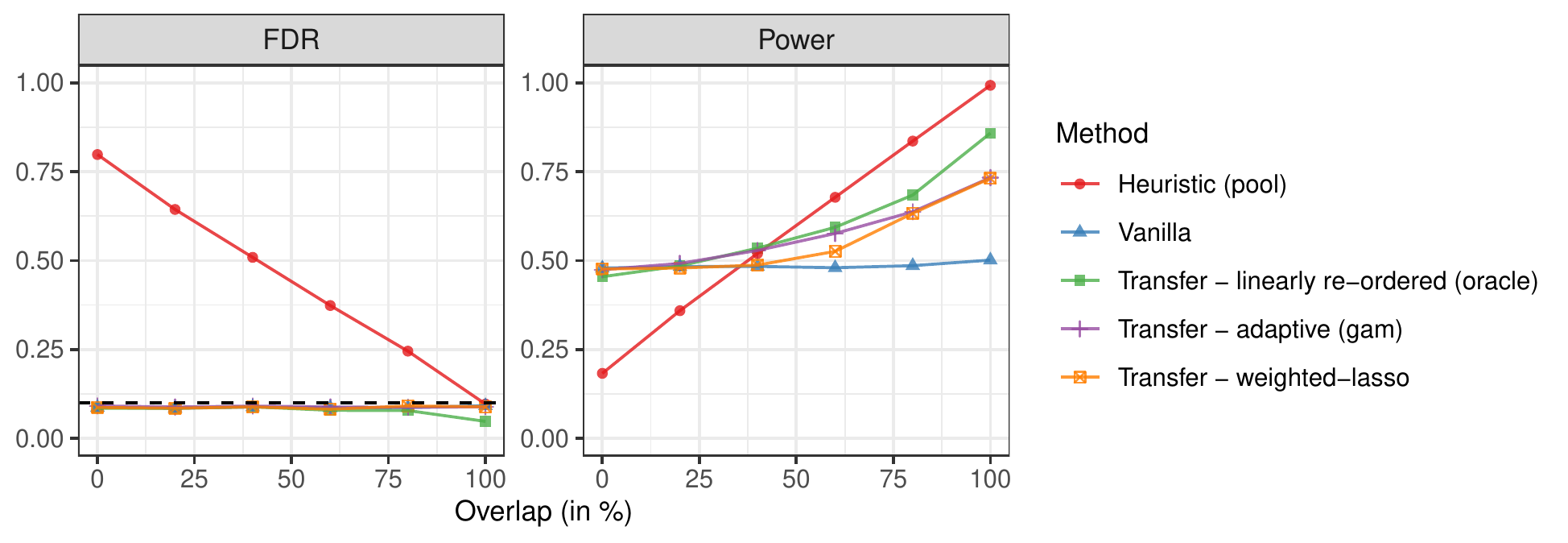}
 \caption{Performance of different transfer learning methods on simulated data, compared to two benchmarks. 
Each point averages the results of 500 independent experiments. }
 \label{fig:transfer_a}
\end{figure}

Figure~\ref{fig:transfer_a} compares the performance of the three transfer learning methods and of the two benchmarks as a function of the overlap between $S^0$ and $S^1$.
The results show that transfer learning helps increase power compared to the vanilla method of \citet{candes2018panning} if the external environments are sufficiently similar to the target one, while always controlling the FDR both when predicted by the theory, and when adopting the oracle method that does not have rigorous guarantees. 
In the case of the adaptive knockoff filter and of the weighted-lasso statistics, transfer learning does not seem to hurt power even if the target environment is completely different from the others (zero overlap). 
In contrast, the method with linearly re-ordered statistics suffers from lower power if the target environment is very different. The power loss is more apparent if $\theta$ is fixed (see Figure~\ref{fig:transfer_b} in Appendix~\ref{app:numerical-exp}). 
 Unsurprisingly, pooling does not control the false discovery rate, as it tends to report any variables that are non-null in at least one environment, not necessarily the target one.
Finally, Figure~\ref{fig:transfer_b} in Appendix~\ref{app:numerical-exp} shows the performance of knockoffs with linearly re-ordered knockoffs statistics applied with different fixed choices of $\theta$. These results suggest that larger values of $\theta$ make the power more sensitive to overlap between $S^0$ and the other environments: if overlap is high, the power is significantly larger than that of the vanillas knockoffs, but if the overlap is low, the power becomes much lower.

\FloatBarrier

\subsection{Real genotype data and simulated phenotypes}
\label{sec:half_synthetic}

The performances of all methods and benchmarks from the previous section are now compared on simulated but realistic GWAS data.
For this purpose, we utilize the genetic data for the UK Biobank \citep{bycroft2018} samples with self-reported ancestry in one of the following five distinct populations: African ($n=7,635$), Asian ($n=3,284$), British ($n = 429,934$), non-British European ($n=28,994$), and Indian ($n=7,628$), for a total of $477,475$ individuals, as in~\citet{li2021searching}.
For each of these individuals, we focus on the variants from chromosome one, disregarding the other chromosomes for simplicity. 
Following the same approach of~\citet{sesia2020controlling}, we only analyze biallelic single nucleotide polymorphisms (SNPs) with minor allele frequency above 0.1\% and in Hardy-Weinberg equilibrium (p-value above $10^{-6}$) among the subset of 350,119 unrelated British individuals previously analyzed in~\citet{sesia2020}.

The genotyped variants are partitioned into contiguous blocks in high linkage disequilibrium, so that their median width is 208 kb.
These blocks are obtained by applying complete-linkage hierarchical clustering using the genetic distances, which are measured in centimorgan and estimated in a European population~\citep{international2010integrating}; see~\citet{sesia2020controlling} for additional details.
As in previous work on knockoffs for GWAS~\citep{sesia2020}
our goal is to discover which of the above SNP blocks are likely to contain distinct association signals, focusing of course on the analysis of the data from the target environment. The SNP blocks mentioned above act as the fundamental units of inference in our analysis, so that the fundamental conditional hypotheses $H_j$ defined in~\eqref{eqn:hypo_e} are effectively replaced with the slightly more general 
\begin{align} \label{eq:null-ci-grouped}
 \mathcal{H}_{G} : Y \indep X_{G} \mid X_{-G},
\end{align}
where $G \subseteq \{1,\ldots,p\}$ indicates a block of SNPs, leaving however all other aspects of the analysis unchanged compared to Section~\ref{section:methods}.
The advantage of this approach is that it allows us to control the trade-off between power and resolution, as the conditional hypotheses corresponding to smaller SNP blocks are naturally more informative, as they allow us to localize the significant genetic associations more precisely, but they are also inevitably more difficult to reject~\citep{sesia2020}.
In practice, we utilize in these experiments the same groups of SNPs and the corresponding knockoffs at the 208kb resolution as in \citet{sesia2020controlling}.

The true model for the phenotype is determined by randomly picking $100$ SNPs as the ``causal variants'', ensuring that they all come from distinct SNP blocks, 
and varying as a control parameter the heterogeneity across populations of their minor allele frequencies.
When the heterogeneity parameter is 0\%, all causal variants have approximately equal minor allele frequencies in all five populations. When the heterogeneity parameter is 100\%, each of the five populations is assigned 20 specific causal variants among those with the highest frequency in that population and lowest possible frequency in all others, consistently with the constraint that each block should contain at most one causal variant. In the intermediate cases in which the heterogeneity parameter is between 0\% and 100\%, a corresponding fraction of causal variants are chosen with the first method above while the remaining ones are chosen with the second method, interpolating between those two extreme approaches.
Note that all 100 causal variants may be present in all populations, although with different frequencies, and have a causal effect in all of them.
The causal effect sizes are however population-specific. In particular, for each population, the effect sizes of all 100 causal variants are independent and identically distributed within the interval $[0.1,10]$. The signs of the causal effects are independent coin flips but remain constant across populations.
Conditional on the genotypes and on the above causal effects, the synthetic phenotypes are generated from a linear model with homoscedastic Gaussian noise:
$Y^e \sim \mathcal{N}(X^e\beta^e, \sigma^2)$.
Above, $X^e$ and $Y^e$ indicate the genotypes and phenotypes in the $e$-th population, respectively, while $\beta^e$ is the vector containing the signed effect sizes for all 100 causal variants. The noise variance $\sigma^2$ is fixed so that the signal-to-noise ratio is 5\% or 10\%, depending on the setting.
The goal of this analysis is to localize the 100 blocks of SNPs containing causal variants controlling the FDR below 10\%.

\begin{figure}[!htb]
 \centering
 \includegraphics[width=\linewidth]{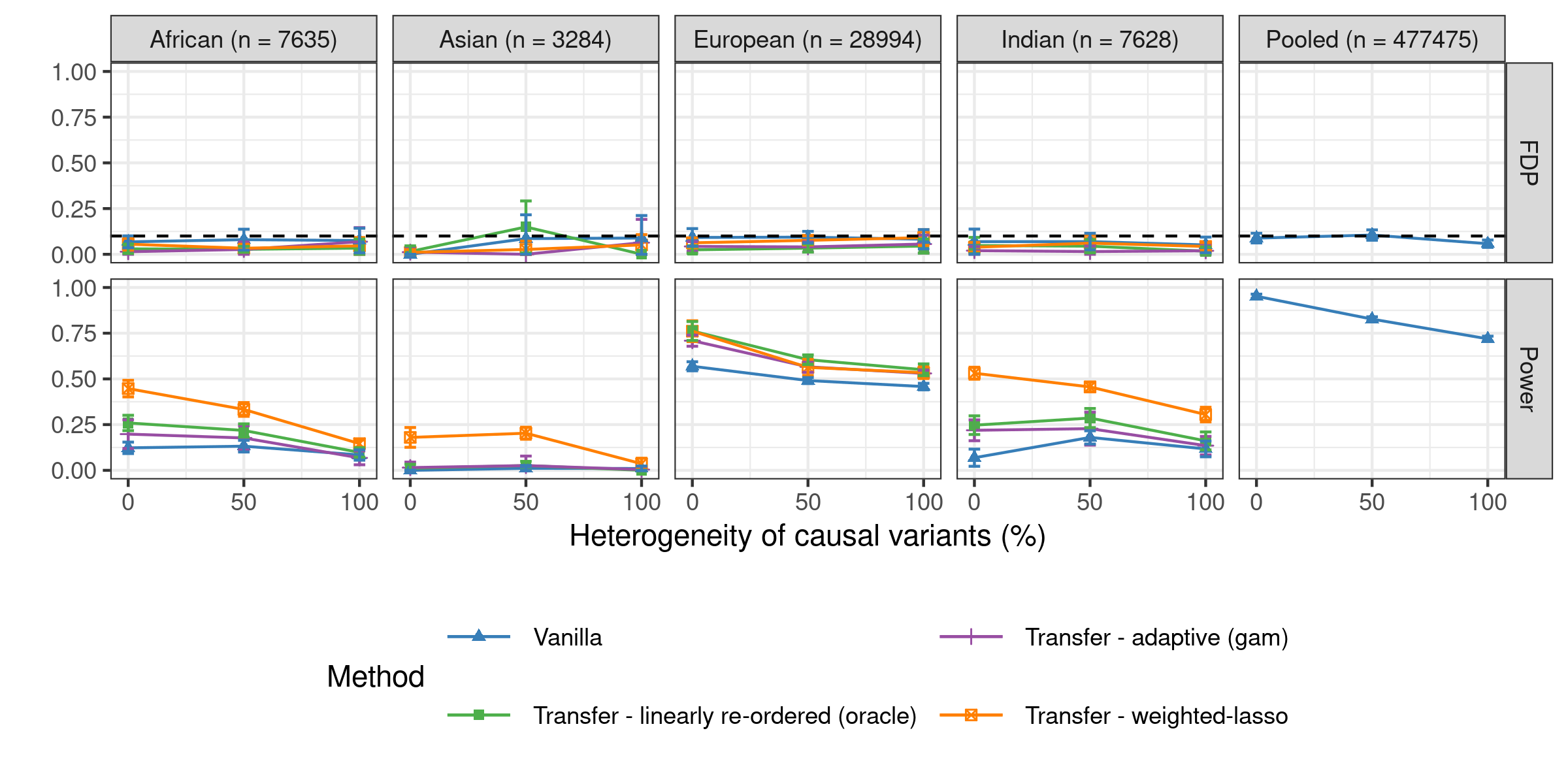}
 \caption{Performances of different knockoff methods for transfer learning applied to simulated GWAS data with real genotypes from different populations. The nominal FDR level is 10\%. The empirical FDR and power are averaged over 10 experiments with independent phenotypes.  The last column summarizes the results of a standard knockoff analysis of the pooled data from all populations.}
 \label{fig:sim_ukb}
\end{figure}

Figure~\ref{fig:sim_ukb} compares the performances of all methods applied to the above data as a function of the heterogeneity of the causal variants, in the case of 10\% signal-to-noise-ratio. Each method is applied to the data from the population labeled in the corresponding column, while the transfer learning prior information is obtained by applying knockoffs to the pooled data set of all UK Biobank individuals---note that this is a valid solution because the support of the causal model is the constant across populations. The results demonstrate that all transfer learning methods control the FDR, as predicted by the theory, but the weighted lasso approach is the most powerful.
Note that the sample sizes for these four minority groups differ meaningfully, largely explaining the variance in the performances of all methods across populations.
However, these groups also differ in their genetic similarity to the British population, and hence in the amount of transferable information, which means that one should not expect transfer learning to perform equally well even if the sample sizes were all the same.
Indeed, Figure~\ref{fig:sim_ukb_small} in Appendix~\ref{app:numerical-exp} reports analogous results obtained from the analysis of smaller data sets with equal sample sizes across populations ($n=3284$, the number of individuals belonging to the Asian population, which is the smallest environment in this data set), showing that transfer learning from the British population tends to be most effective when the target population is the European one, especially if the heterogeneity of the causal allele frequencies is high.

\begin{figure}[!htb]
 \centering
 \includegraphics[width=\linewidth]{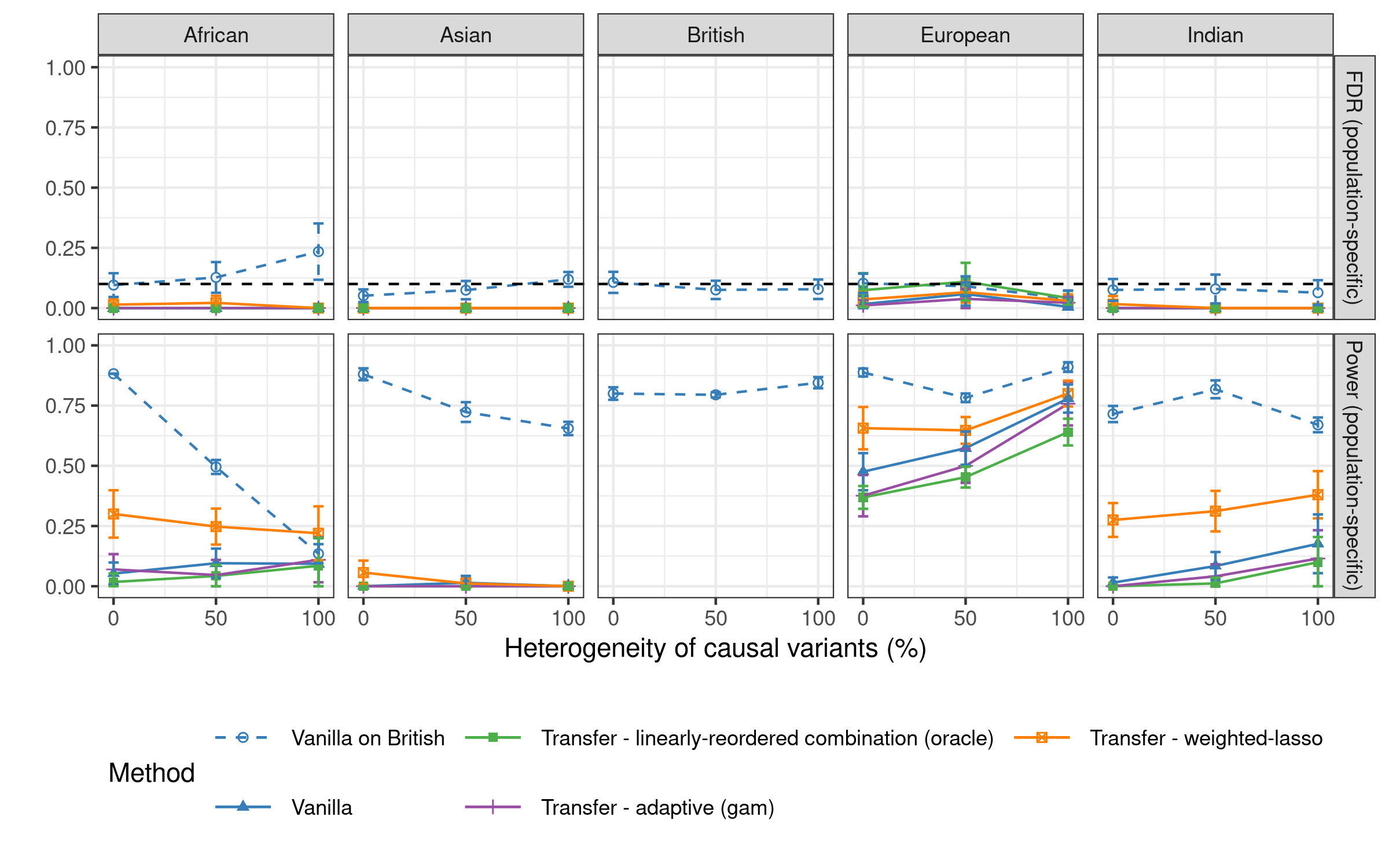}
 \caption{Performances of different knockoff methods for transfer learning applied to simulated GWAS data, as in Figure~\ref{fig:sim_ukb}. The FDR and power are population-specific: they only count discoveries whose minor allele frequency is highest in the population of interest.}
 \label{fig:sim_ukb_specific}
\end{figure}

Figure~\ref{fig:sim_ukb_specific} compares the performances of the different transfer learning methods evaluated in a population-specific sense, when the signal-to-noise ratio is 5\%.
More precisely, we imagine assigning each variant, whether causal or not, to the population in which it displays the largest minor allele frequency. Then, the FDR and power of our analysis are evaluated separately in each population by counting only the specific discoveries and true causal variants assigned to it. 
The intuition is that these are the specific discoveries that will likely be most useful for building effective predictive models for the population of interest.
Unfortunately, controlling the false discovery rate becomes difficult in this setting, as some of the reported associations are discarded post-hoc, because one generally has no guarantee that the expected spurious findings are not disproportionately concentrated among the selected ones \citep{katsevich2018controlling}.
Although the transfer learning methods described in this paper are not the only possible approach to mitigate this problem, and in truth they do not even explicitly solve the post-hoc filtering challenge explained above, we will demonstrate that they can be useful to highlight discoveries that are significant both statistically and practically within the target environment.
The results show that all transfer learning methods considered here always control the population-specific FDR in practice, although they are only theoretically guaranteed to control the global FDR over all reported variants. By contrast, the vanilla knockoffs method applied to the British samples alone does not always empirically control the population-specific FDR in other populations, even though one may have intuitively expected it to be valid also in this sense since the causal model is by design the same regardless of ancestry. However, the issue with the vanilla single-population analysis here is that there is relatively little information in the British samples about the variants which are specific to other groups, and so the knockoff filter is more likely to make mistakes on them.
Figure~\ref{fig:sim_ukb_specific_10} in Appendix~\ref{app:numerical-exp} shows similar results in the case of signal-to-noise-ratio equal to 10\%; the population-specific FDR violation of the vanilla approach is still noticeable here but it is smaller and not statistically significant.
Finally, Figure~\ref{fig:sim_ukb_specific_pooled} shows that pooling appears to empirically control the population-specific FDR in the setting of Figure~\ref{fig:sim_ukb_specific}, even though the vanilla approach applied to British samples did not.

\FloatBarrier

\section{Real data analysis}
%\color{blue}
%? comparison with other methods for transfer learning Ex. weighted FDR control following Katryn R, Edgard;  adapT -- which would be closer to Zhimei's method.
%\color{black}

We apply the different transfer learning methods discussed in this paper to study several real phenotypes in the UK Biobank resource, using the same data and knockoffs as in \citet{sesia2020controlling}.
Following the approach of Section~\ref{sec:half_synthetic}, we consider five environments based on 
the self-reported ancestry of each individual, separately utilizing as target environments each of the four minority populations: African, Asian, non-British European, and Indian.
In each case, the test statistics computed on the British samples are used as prior information. 
In particular, we consider two types
of prior information: single-phenotype prior information and multi-phenotype prior
information. The former consists of the importance statistics obtained from the analysis of the phenotype of interest in the British population, 
while the latter includes also the analogous information obtained from the analysis of other related phenotypes with potentially similar genetic architectures.
All data are pre-processed based on the protocol of \citet{sesia2020controlling}, which consists of filtering SNPs and individuals based on standard quality-control criteria, clustering the genotypes into hierarchical blocks at different resolutions (with median widths equal to 3, 20, 41, 81, 208, 425 kb, respectively) following the same approach described in Section~\ref{sec:half_synthetic},
and generating the corresponding knockoffs at each resolution.
For all methods, the knockoff filter (SeqStep) of \citet{barber2015controlling} is applied separately at each resolution~\citep{sesia2020}. The offset parameter of this filter is set equal to $0$, which tends to be more powerful than the standard implementation but does not theoretically control the FDR if the number of discoveries is very small.

\subsection{Leveraging single-phenotype prior information}

We analyze here three phenotypes: platelet count, standing height, and body mass index (BMI). 
For each phenotype, we make use of the feature importance statistic $W$ corresponding to the same phenotype
obtained from the British population. The three transfer learning methods introduced 
in Section~\ref{section:methods} are implemented and compared with the results of the 
vanilla knockoffs procedure. 
Table~\ref{tab:platelet} summarizes the numbers of discoveries for platelet count.
The results show that transfer learning yields more discoveries than vanilla knockoffs, confirming the advantage of leveraging
prior information. The knockoffs procedure with weighted lasso statistics leads to the most discoveries, demonstrating the benefit of refitting the predictive model.
Analogous results, although with fewer discoveries, are shown in Tables~\ref{tab:height}-\ref{tab:bmi} (Appendix~\ref{app:numerical-data}) for height and BMI, respectively.
The full lists of discoveries are available online from \url{https://msesia.github.io/knockoffgwas/ukbiobank}.

\begin{table}[ht]
\centering
\begin{tabular}{l|c|cccc}
\toprule
Population & Resolution (kb) & \makecell{Vanilla}
           & \makecell{Transfer -\\ linearly re-ordered \\ ($\theta=0.1$)} 
           & \makecell{Transfer -\\adaptive (gam) } & \makecell{Transfer -\\ weighted-lasso}\\
\midrule
 & 3 & 4 & 6 & 5 & 26\\
%\cmidrule{2-6}
 & 20 & 27 & 31 & 53 & 55\\
%\cmidrule{2-6}
 & 41 & 56 & 65 & 91 & 112\\
%\cmidrule{2-6}
 & 81 & 70 & 81 & 48 & 190\\
%\cmidrule{2-6}
 & 208 & 61 & 74 & 163 & 306\\
%\cmidrule{2-6}
\multirow{-6}{*}{\raggedright\arraybackslash European} & 425 & 76 & 100 & 151 & 320\\
\cmidrule{1-6}
 & 3 & 1 & 1 & 1 & 1\\
%\cmidrule{2-6}
 & 20 & 6 & 6 & 5 & 9\\
%\cmidrule{2-6}
 & 41 & 3 & 3 & 9 & 6\\
%\cmidrule{2-6}
 & 81 & 4 & 13 & 16 & 34\\
%\cmidrule{2-6}
 & 208 & 20 & 28 & 24 & 49\\
%\cmidrule{2-6}
\multirow{-6}{*}{\raggedright\arraybackslash Indian} & 425 & 22 & 31 & 13 & 63\\
\cmidrule{1-6}
 & 3 & 4 & 5 & 7 & 11\\
%\cmidrule{2-6}
 & 20 & 5 & 6 & 5 & 6\\
%\cmidrule{2-6}
 & 41 & 11 & 2 & 12 & 20\\
%\cmidrule{2-6}
 & 81 & 16 & 23 & 18 & 27\\
%\cmidrule{2-6}
 & 208 & 12 & 24 & 22 & 27\\
%\cmidrule{2-6}
\multirow{-6}{*}{\raggedright\arraybackslash African} & 425 & 10 & 17 & 12 & 12\\
\cmidrule{1-6}
 & 3 & 4 & 5 & 6 & 4\\
%\cmidrule{2-6}
 & 20 & 4 & 7 & 5 & 1\\
%\cmidrule{2-6}
 & 41 & 7 & 6 & 7 & 5\\
%\cmidrule{2-6}
 & 81 & 3 & 5 & 7 & 7\\
%\cmidrule{2-6}
 & 208 & 0 & 0 & 0 & 4\\
%\cmidrule{2-6}
\multirow{-6}{*}{\raggedright\arraybackslash Asian} & 425 & 0 & 0 & 0 & 12\\
\bottomrule
\end{tabular}
\caption{Numbers of discoveries for platelet count reported 
by vanilla knockoffs and three transfer 
learning methods applied to data from different minority 
populations in the UK Biobank, leveraging prior information 
from a much larger number of British samples. 
The discoveries are reported separately at six different levels of 
resolution. The nominal FDR level is 10\% at each resolution.
}
\label{tab:platelet}
\end{table}

\subsection{Leveraging multi-phenotype prior information} \label{eq:multi-pheno}

We now consider four related phenotypes that may share some similarities in their genetic architectures: BMI, systolic blood pressure (SBP),
diabetes, and cardiovascular disease (CVD). 
These four phenotypes are separately studied using the British samples, and the results of all those analyses are utilized jointly as prior information for all specific studies in the minority populations.
Concretely, before studying any phenotype in the minority populations, we first compute on the data from the British samples the $2p$-dimensional estimated 
coefficients $\hat{b}^{\text{diabetes}}, \hat{b}^{\text{BMI}},\hat{b}^{\text{CVD}},
\hat{b}^{\text{SBP}}$---and the corresponding knockoff test statistics
$W^{\text{diabetes}}, W^{\text{BMI}}, W^{\text{CVD}}$, $W^{\text{SBP}}$---for diabetes, BMI, CVD, and SBP, respectively.
Then, the adaptive knockoffs method is simply applied to each minority study with the four-dimensional vector $(W_j^{\text{diabetes}},
W_j^{\text{BMI}}, W_j^{\text{CVD}}, W_j^{\text{SBP}})$ as prior information input \citep{ren2020knockoffs} for each SNP $j \in \{1,\ldots,p\}$,
and utilizing the single-source weighted lasso statistics as the test statistic.
In the case of transfer learning with weighted-lasso statistics, which require one-dimensional prior information for each SNP, we combine $(\hat{b}_j^{\text{diabetes}}$, $\hat{b}_j^{\text{BMI}}$, $\hat{b}_j^{\text{CVD}}, \hat{b}_j^{\text{SBP}})$ by taking the mean and then proceed as in Section~\ref{subsection:penalty}; we will henceforth
refer to this method as the knockoffs procedure with multi-weighted lasso statistics.

Table~\ref{tab:multi_bmi_small} reports the number of findings for BMI obtained from the analysis of the European samples. These results show that transfer learning increases power, consistently with the numerical experiments in Section~\ref{sec:experiments}.
Tables~\ref{tab:multi_diabetes}-\ref{tab:multi_cvd} in Appendix~\ref{app:numerical-exp} summarize the analogous numbers of discoveries corresponding to the other minority populations and phenotypes. Again, one can see that transfer learning tends to lead to some benefits, although for these phenotypes the power of all methods is often very low in all but the European population, likely due to the relatively small sample sizes.
In this example, taking the average of the prior information
from difference sources does not improve the performance
of transfer learning with weighted lasso statistics, possibly because it is not
optimal to assign an equal weight to each source. The adaptive 
knockoffs methodology offers a more principled and tuning-free approach for dealing with 
multi-dimensional prior information, and in our analysis, we see 
improvement in some cases. Since here both the power and improvement
are small, we do not attempt to draw conclusions on the performance 
of the methods, but rather try to demonstrate the possibility of simultaneously
leveraging prior information from multiple sources which differ both in their underlying populations and in the phenotypes studied.

\begin{table}[!htb]
\begin{tabular}[t]{l|c|cccc}
\toprule
Population & Resolution & Vanilla 
           & \makecell{Transfer -\\ adaptive (gam)} 
           & \makecell{Transfer -\\ weighted-lasso}
           & \makecell{Transfer -\\ multi-weighted-lasso}\\
\midrule
 & 3 & 1 & 1 & 0 & 0\\
%\cmidrule{2-6}
 & 20 & 6 & 7 & 7 & 8\\
%\cmidrule{2-6}
 & 41 & 7 & 13 & 11 & 10\\
%\cmidrule{2-6}
 & 81 & 3 & 2 & 21 & 2\\
%\cmidrule{2-6}
 & 208 & 9 & 16 & 25 & 12\\
%\cmidrule{2-6}
\multirow{-6}{*}{\raggedright\arraybackslash European} & 425 & 7 & 50 & 47 & 22\\
\bottomrule
\end{tabular}
\caption{Numbers of discoveries for BMI, using UK Biobank data from non-British European samples
and leveraging prior knowledge on BMI, systolic blood pressure,
diabetes, and cardiovascular disease acquired from the analysis of British samples. The discoveries are reported at six different levels of resolution. The nominal FDR level is 10\% at each resolution.}
\label{tab:multi_bmi_small}
\end{table}

\section{Discussion}

This paper demonstrated that incorporating relevant knowledge from external data sets can significantly improve the power of conditional testing with knockoffs, especially if the available prior information is carefully leveraged to fit a more accurate predictive model for the computation of the test statistics.
The empirical results obtained on synthetic and real data suggest that these methods can be useful for the analysis of GWAS data, allowing one to increase the number of discoveries by borrowing strength from possibly larger studies which may either involve populations with different ancestries or focus on different phenotypes. Transfer learning is indeed particularly appealing in such cases because the naive alternative of pooling all the data would not be a satisfactory solution, either because it may skew our findings towards those which are most relevant for the most common population, or because it would not lead to the discovery of the variables that are truly important for the outcome of interest.

An intriguing direction for future research would involve investigating possible connections between the transfer learning for {\em conditional testing} studied in this paper, and the more traditional task of {\em predictive transfer learning}, with the ultimate goal of computing more accurate estimates of genetic risk for populations which have been historically underrepresented in GWAS data. In fact, one may intuitively expect that predictive models based on variables selected with the methods in this paper may perform better than those obtained with alternative variable selection approaches that either do not leverage relevant information gained by analyzing data from different populations, or do not correctly account for genetic diversity across populations.

Regarding future applications to GWAS data, it is possible to apply the transfer learning methods in Section~\ref{eq:multi-pheno} to leverage prior information acquired from the study of related phenotypes within the same population as the target one (e.g., the British population in the UK Biobank). One must though be careful to avoid utilizing data from the same individuals twice, as our theoretical results assume the prior information to be independent of the test statistics computed on the target data set. Since the sampling randomness upon which knockoff inferences are based arises from the distribution of the genotypes, not that of the phenotypes, our transfer learning methods would not be guarantee to control the FDR if the prior depended on observations involving individuals also present in the target data set, regardless of the relation between the different phenotypes. However, it would be correct to utilize two different subsets of British samples to acquire potentially useful prior information and to calculate the test statistics for the phenotype of interest. This may be particularly relevant if the phenotype is only measured in relatively few people or, in the case of a binary trait, there are many more healthy controls than disease cases. Then, transfer learning would be a natural solution to gain strength from data on more common conditions suspected to share some similarities in genetic architectures, while retaining guaranteed FDR control for the study of interest. 

\section*{Acknowledgements}

We gratefully acknowledge support from NIH grants R56HG010812, R01MH113078 and R01MH123157.
Z.R.~acknowledges the support from ONR grant N00014-20-1-2337.
The authors thank the Research Computing Center at Stanford University for computing resources, as well as 
the participants and investigators of the UK Biobank (application 27837). 

\FloatBarrier

\printbibliography

@article{international2010integrating,
  title={Integrating common and rare genetic variation in diverse human populations},
  author={International HapMap 3 Consortium and others},
  journal={Nature},
  volume={467},
  number={7311},
  pages={52},
  year={2010},
  publisher={Europe PMC Funders}
}

@article{sesia2020,
author={Sesia, Matteo and Katsevich, Eugene and Bates, Stephen and Cand{\`e}s, Emmanuel and Sabatti, Chiara},
title={Multi-resolution localization of causal variants across the genome},
journal={Nat. Comm.},
year={2020},
volume={11},
number={1},
pages={1093},
}

@article{laan1997demographic,
  title={Demographic history and linkage disequilibrium in human populations},
  author={Laan, Maris and P{\"a}{\"a}bo, Svante},
  journal={Nat. Genet.},
  volume={17},
  number={4},
  pages={435--438},
  year={1997},
  publisher={Nature Publishing Group}
}

@article{benjamini2008screening,
  title={Screening for partial conjunction hypotheses},
  author={Benjamini, Yoav and Heller, Ruth},
  journal={Biometrics},
  volume={64},
  number={4},
  pages={1215--1222},
  year={2008},
  publisher={Wiley Online Library}
}

@article{sirugo2019missing,
  title={The missing diversity in human genetic studies},
  author={Sirugo, Giorgio and Williams, Scott M and Tishkoff, Sarah A},
  journal={Cell},
  volume={177},
  number={1},
  pages={26--31},
  year={2019},
  publisher={Elsevier}
}

@Article{wray2013,
   Author="Wray, N. R.  and Yang, J.  and Hayes, B. J.  and Price, A. L.  and Goddard, M. E.  and Visscher, P. M. ",
   Title="{{P}itfalls of predicting complex traits from {S}{N}{P}s}",
   Journal="Nat Rev Genet",
   Year="2013",
   Volume="14",
   Number="7",
   Pages="507--515",
   Month="07"
}

@Article{popejoy2016,
   Author="Popejoy, A. B.  and Fullerton, S. M. ",
   Title="{{G}enomics is failing on diversity}",
   Journal="Nature",
   Year="2016",
   Volume="538",
   Number="7624",
   Pages="161--164",
   Month="10"
}

@Article{price2006,
   Author="Price, A. L.  and Patterson, N. J.  and Plenge, R. M.  and Weinblatt, M. E.  and Shadick, N. A.  and Reich, D. ",
   Title="{{P}rincipal components analysis corrects for stratification in genome-wide association studies}",
   Journal="Nat Genet",
   Year="2006",
   Volume="38",
   Number="8",
   Pages="904--909",
   Month="Aug"
}

@Article{rosenberg2002,
   Author="Rosenberg, N. A.  and Pritchard, J. K.  and Weber, J. L.  and Cann, H. M.  and Kidd, K. K.  and Zhivotovsky, L. A.  and Feldman, M. W. ",
   Title="{{G}enetic structure of human populations}",
   Journal="Science",
   Year="2002",
   Volume="298",
   Number="5602",
   Pages="2381--2385",
   Month="Dec"
}

@Article{Reich2018,
   Author="Reich, D.",
   Title="{How Genetics Is Changing Our Understanding of `Race'}",
   Journal="The New York Times",
   Year="2018",
   Volume="",
   Month="March 23"
}

@Article{holmes2018,
   Author="Holmes, I.",
   Title="{{W}hat Happens When Geneticists Talk Sloppily About Race}",
   Journal="The Atlantic",
   Year="2018",
   Volume="",
   Month="April 25"
}

@Article{martin2017,
   Author="Martin, A. R.  and Gignoux, C. R.  and Walters, R. K.  and Wojcik, G. L.  and Neale, B. M.  and Gravel, S.  and Daly, M. J.  and Bustamante, C. D.  and Kenny, E. E. ",
   Title="{{H}uman {D}emographic {H}istory {I}mpacts {G}enetic {R}isk {P}rediction across {D}iverse {P}opulations}",
   Journal="Am J Hum Genet",
   Year="2017",
   Volume="100",
   Number="4",
   Pages="635--649",
   Month="Apr"
}

@Article{martin2019,
   Author="Martin, A. R.  and Kanai, M.  and Kamatani, Y.  and Okada, Y.  and Neale, B. M.  and Daly, M. J. ",
   Title="{{C}linical use of current polygenic risk scores may exacerbate health disparities}",
   Journal="Nat Genet",
   Year="2019",
   Volume="51",
   Number="4",
   Pages="584--591",
   Month="04"
}

@Article{bycroft2018,
    author={Bycroft, Clare and Freeman, Colin and Petkova, Desislava and Band, Gavin and Elliott, Lloyd T. and Sharp, Kevin and Motyer, Allan and Vukcevic, Damjan and Delaneau, Olivier and O'Connell, Jared and Cortes, Adrian and Welsh, Samantha and Young, Alan and Effingham, Mark and McVean, Gil and Leslie, Stephen and Allen, Naomi and Donnelly, Peter and Marchini, Jonathan},
    title = {The {UK} Biobank resource with deep phenotyping and genomic data},
    journal = {Nature},
    volume = {562},
    pages = {203--209},
    year = 2018,
}

@article{sesia2018,
    author = {Sesia, M and Sabatti, C and Cand{\`e}s, E},
    title = "{Gene hunting with hidden Markov model knockoffs}",
    journal = {Biometrika},
    volume = {106},
    number = {1},
    pages = {1-18},
    year = {2018},
    month = {08},
    issn = {0006-3444},
    doi = {10.1093/biomet/asy033},
    url = {https://doi.org/10.1093/biomet/asy033},
    eprint = {https://academic.oup.com/biomet/article-pdf/106/1/1/27774559/asy033.pdf},
}

@article{sesia2020controlling,
	author = {Sesia, Matteo and Bates, Stephen and Cand{\`e}s, Emmanuel and Marchini, Jonathan and Sabatti, Chiara},
	title = {{FDR} control in {GWAS} with population structure},
	elocation-id = {2020.08.04.236703},
	year = {2021},
	publisher = {Cold Spring Harbor Laboratory},
	journal = {preprint at bioRxiv}
}

@article{heckman1979sample,
  title={Sample selection bias as a specification error},
  author={Heckman, James J},
  journal={Econometrica},
  pages={153--161},
  year={1979},
  publisher={JSTOR}
}

@article{pan2009survey,
  title={A survey on transfer learning},
  author={Pan, Sinno Jialin and Yang, Qiang},
  journal={IEEE Transactions on knowledge and data engineering},
  volume={22},
  number={10},
  pages={1345--1359},
  year={2009},
  publisher={IEEE}
}

@Article{Duncan2019,
   Author="Duncan, L.  and Shen, H.  and Gelaye, B.  and Meijsen, J.  and Ressler, K.  and Feldman, M.  and Peterson, R.  and Domingue, B. ",
   Title="{{A}nalysis of polygenic risk score usage and performance in diverse human populations}",
   Journal="Nat. Commun.",
   Year="2019",
   Volume="10",
   Number="1",
   Pages="3328",
   Month="07"
}

@article{katsevich2018controlling,
  title={Filtering the rejection set while preserving false discovery rate control},
  author={Katsevich, Eugene and Sabatti, Chiara and Bogomolov, Marina},
  journal={J. Am. Stat. Assoc},
  number={just-accepted},
  pages={1--27},
  year={2021},
  publisher={Taylor \& Francis}
}

@article{li2021searching,
  title={Searching for consistent associations with a multi-environment knockoff filter},
  author={Li, Shuangning and Sesia, Matteo and Romano, Yaniv and Cand{\`e}s, Emmanuel and Sabatti, Chiara},
  journal={arXiv preprint arXiv:2106.04118},
  year={2021}
}

@Article{coram2017,
   Author="Coram, M. A.  and Fang, H.  and Candille, S. I.  and Assimes, T. L.  and Tang, H. ",
   Title="{{L}everaging {M}ulti-ethnic {E}vidence for {R}isk {A}ssessment of {Q}uantitative {T}raits in {M}inority {P}opulations}",
   Journal="Am J Hum Genet",
   Year="2017",
   Volume="101",
   Number="2",
   Pages="218--226",
   Month="Aug"
}

@Article{wojcik2019,
   Author="Wojcik, G. L.  and Graff, M.  and Nishimura, K. K.  and Tao, R.  and Haessler, J.  and Gignoux, C. R.  and Highland, H. M.  and Patel, Y. M.  and Sorokin, E. P.  and Avery, C. L.  and Belbin, G. M.  and Bien, S. A.  and Cheng, I.  and Cullina, S.  and Hodonsky, C. J.  and Hu, Y.  and Huckins, L. M.  and Jeff, J.  and Justice, A. E.  and Kocarnik, J. M.  and Lim, U.  and Lin, B. M.  and Lu, Y.  and Nelson, S. C.  and Park, S. L.  and Poisner, H.  and Preuss, M. H.  and Richard, M. A.  and Schurmann, C.  and Setiawan, V. W.  and Sockell, A.  and Vahi, K.  and Verbanck, M.  and Vishnu, A.  and Walker, R. W.  and Young, K. L.  and Zubair, N.  and Acuña-Alonso, V.  and Ambite, J. L.  and Barnes, K. C.  and Boerwinkle, E.  and Bottinger, E. P.  and Bustamante, C. D.  and Caberto, C.  and Canizales-Quinteros, S.  and Conomos, M. P.  and Deelman, E.  and Do, R.  and Doheny, K.  and Fernández-Rhodes, L.  and Fornage, M.  and Hailu, B.  and Heiss, G.  and Henn, B. M.  and Hindorff, L. A.  and Jackson, R. D.  and Laurie, C. A.  and Laurie, C. C.  and Li, Y.  and Lin, D. Y.  and Moreno-Estrada, A.  and Nadkarni, G.  and Norman, P. J.  and Pooler, L. C.  and Reiner, A. P.  and Romm, J.  and Sabatti, C.  and Sandoval, K.  and Sheng, X.  and Stahl, E. A.  and Stram, D. O.  and Thornton, T. A.  and Wassel, C. L.  and Wilkens, L. R.  and Winkler, C. A.  and Yoneyama, S.  and Buyske, S.  and Haiman, C. A.  and Kooperberg, C.  and Le Marchand, L.  and Loos, R. J. F.  and Matise, T. C.  and North, K. E.  and Peters, U.  and Kenny, E. E.  and Carlson, C. S. ",
   Title="{{G}enetic analyses of diverse populations improves discovery for complex traits}",
   Journal="Nature",
   Year="2019",
   Volume="570",
   Number="7762",
   Pages="514--518",
   Month="06"
}

@article{candes2018panning,
  title={Panning for gold: 'model-{X}' knockoffs for high dimensional controlled variable selection},
  author={Cand{\`e}s, Emmanuel and Fan, Yingying and Janson, Lucas and Lv, Jinchi},
  journal={Journal of the Royal Statistical Society: Series B (Statistical Methodology)},
  volume={80},
  number={3},
  pages={551--577},
  year={2018},
  publisher={Wiley Online Library}
}

@article{ren2020knockoffs,
  title={Knockoffs with side information},
  author={Ren, Zhimei and Cand{\`e}s, Emmanuel},
  journal={arXiv preprint arXiv:2001.07835},
  year={2020}
}

@article{sesia2019gene,
  title={Gene hunting with hidden Markov model knockoffs},
  author={Sesia, Matteo and Sabatti, Chiara and Cand{\`e}s, Emmanuel J},
  journal={Biometrika},
  volume={106},
  number={1},
  pages={1--18},
  year={2019},
  publisher={Oxford University Press}
}

@article{romano2020deep,
  title={Deep knockoffs},
  author={Romano, Yaniv and Sesia, Matteo and Cand{\`e}s, Emmanuel},
  journal={Journal of the American Statistical Association},
  volume={115},
  number={532},
  pages={1861--1872},
  year={2020},
  publisher={Taylor \& Francis}
}

@article{barber2015controlling,
  title={Controlling the false discovery rate via knockoffs},
  author={Barber, Rina Foygel and Cand{\`e}s, Emmanuel},
  journal={Ann. Stat.},
  volume={43},
  number={5},
  pages={2055--2085},
  year={2015},
  publisher={Institute of Mathematical Statistics}
}

@article{benjamini1995,
    author = {Benjamini, Y. and Hochberg, Y.},
    title = {Controlling the false discovery rate: a practical and powerful approach to multiple testing},
    journal = {J. R. Stat. Soc. B.},
    volume = {57},
    pages = {289--300},
    year = 1995,
}

@article{genovese2006false,
  title={False discovery control with p-value weighting},
  author={Genovese, Christopher R and Roeder, Kathryn and Wasserman, Larry},
  journal={Biometrika},
  volume={93},
  number={3},
  pages={509--524},
  year={2006},
  publisher={Oxford University Press}
}

@article{ignatiadis2016data,
  title={Data-driven hypothesis weighting increases detection power in genome-scale multiple testing},
  author={Ignatiadis, Nikolaos and Klaus, Bernd and Zaugg, Judith B and Huber, Wolfgang},
  journal={Nature methods},
  volume={13},
  number={7},
  pages={577--580},
  year={2016},
  publisher={Nature Publishing Group}
}

@article{ignatiadis2017covariate,
  title={Covariate powered cross-weighted multiple testing},
  author={Ignatiadis, Nikolaos and Huber, Wolfgang},
  journal={arXiv preprint arXiv:1701.05179},
  year={2017}
}

@article{lei2018adapt,
  title={AdaPT: an interactive procedure for multiple testing with side information},
  author={Lei, Lihua and Fithian, William},
  journal={Journal of the Royal Statistical Society: Series B (Statistical Methodology)},
  volume={80},
  number={4},
  pages={649--679},
  year={2018},
  publisher={Wiley Online Library}
}

@article{roquain2009optimal,
  title={Optimal weighting for false discovery rate control},
  author={Roquain, Etienne and Van De Wiel, Mark A},
  journal={Electronic journal of statistics},
  volume={3},
  pages={678--711},
  year={2009},
  publisher={Institute of Mathematical Statistics and Bernoulli Society}
}

@article{hu2010false,
  title={False discovery rate control with groups},
  author={Hu, James X and Zhao, Hongyu and Zhou, Harrison H},
  journal={Journal of the American Statistical Association},
  volume={105},
  number={491},
  pages={1215--1227},
  year={2010},
  publisher={Taylor \& Francis}
}

@article{durand2019adaptive,
  title={Adaptive $p$-value weighting with power optimality},
  author={Durand, Guillermo},
  journal={Electronic Journal of Statistics},
  volume={13},
  number={2},
  pages={3336--3385},
  year={2019},
  publisher={Institute of Mathematical Statistics and Bernoulli Society}
}

@article{zhao2014weighted,
  title={Weighted p-value procedures for controlling FDR of grouped hypotheses},
  author={Zhao, Haibing and Zhang, Jiajia},
  journal={Journal of Statistical Planning and Inference},
  volume={151},
  pages={90--106},
  year={2014},
  publisher={Elsevier}
}

\appendix

% Special numbering for the appendix
\setcounter{figure}{0}    
\setcounter{table}{0}    
\setcounter{algorithm}{0}    
\setcounter{equation}{0}    
\renewcommand\thetable{A\arabic{table}}
\renewcommand\thefigure{A\arabic{figure}}
\renewcommand\thealgorithm{A\arabic{algorithm}}
\renewcommand\theequation{A\arabic{equation}}
\renewcommand{\figurename}{Figure}
\renewcommand{\tablename}{Table}
\makeatletter
\renewcommand{\ALG@name}{Algorithm}
\makeatother

\section{Mathematical proofs}
\subsection{Proof of Proposition \ref{prop:coin_lc}}
By Lemma 2 in \citet{candes2018panning}, we know that, conditional on $|W^0|$, the signs of $W^0_j$ are i.i.d.~coin flips for all null $j$ such that $H_j^0$ is true. Since the prior importance statistics $W^{\ext}$ are independent of $W^0$, they are also independent of the signs of $W^0_j$ for all null $j$. 
%Thus, $W^{\lc}$, as a function of $|W^0|$ and $|W^{\ext}|$, is independent of the signs of $W^0_j$ for all null $j \in H_0$. 
%Again by Lemma 2 in \citep{candes2018panning}, the signs of $W^0_j$ are i.i.d.~coin flips for null $j \in H_0$.
Thus, conditional on $|W^0|$ and $|W^{\ext}|$ (and hence also on $|W^{\lc}|$), the signs of $W_j^{\lc}$ for null $j$ are i.i.d.~coin flips.

\subsection{Proof of Proposition \ref{prop:coin_wl}}

The proof follows closely that of Lemma 2 in \citet{candes2018panning}, which did not explicitly consider statistics depending on external data, as we do, but whose main idea is sufficiently general to imply our result easily. 
Let $\varepsilon=(\varepsilon_1, \dots, \varepsilon_p)$ be a sequence of independent random variables such that $\varepsilon_{j}=\pm 1$ with probability $1/2$ if $j$ is null based on $H_{j}^0$, and $\varepsilon_{j}=1$ otherwise. To prove the claim, it suffices to establish that
$
\smash{W \stackrel{d}{=} \varepsilon \odot W}
$,
where the symbol $\odot$ denotes pointwise multiplication, i.e., $\varepsilon \odot W=\left(\varepsilon_{1} W_{1}, \ldots, \varepsilon_{p} W_{p}\right)$. Now, take $\varepsilon$ as above and $S=\left\{j: \varepsilon_{j}=-1\right\} \subseteq \mathcal{H}_{0}$, where $\mathcal{H}_{0}$ denotes the set of null variables in the target environment.
Let $w^{\wl}$ denote the vector-valued function computing the weighted lasso knockoff statistics, i.e., $W^{\wl} = w^{\wl}(\bY^0, [\bX^0, \tilde{\bX}^0], \phi)$. Let $\smash{[\bX^0, \tilde{\bX}^0]_{\text{swap}(S)}}$ be the matrix obtained from $[\bX^0, \tilde{\bX}^0]$ after swapping the column $\bX^0_j$ with the corresponding $\tilde{\bX}^0_j$ for all $j \in S$, and consider $\smash{W^{\wl}_{\text{swap}(S)} }= \smash{w^{\wl}(\bY^0, [\bX^0, \tilde{\bX}^0]_{\text{swap}(S)}, \phi)}$: the statistics computed after swapping the variables in $S$ with their corresponding knockoffs, for all observations in the target environment. 
First, it follows from the flip-sign property that $\smash{W^{\wl}_{\text {swap}(S)}}=\varepsilon \odot W$. Second, Lemma 1 in \citet{candes2018panning} implies that $[\bX^0, \tilde{\bX}^0] \, \, \smash{\stackrel{d}{=}} \, \, [\bX^0, \tilde{\bX}^0]_{\text{swap}(S)} \mid \bY^0$, which, together with the fact that $\phi$ is independent of $([\bX^0, \tilde{\bX}^0], \bY^0)$, further implies that 
$W^{\wl}_{\text {swap }(S)} \, \smash{\stackrel{d}{=}} \, W^{\wl}$. Combining these two results gives the desired property of $\smash{W^{\wl} \stackrel{d}{=} \varepsilon \odot W^{\wl}}$. 

\subsection{Proof of Proposition \ref{prop:coin_wl-2}}

Define $\varepsilon$ and $S$ as in the proof of Proposition \ref{prop:coin_wl}. Then, denote as $w^{\wl}$ the vector-valued function of the data from all environments computing the weighted lasso knockoff statistics, i.e., $W^{\wl} = w^{\wl}(\bY^0, [\bX^0, \tilde{\bX}^0], \bY^{\ext}, [\bX^{\ext}, \tilde{\bX}^{\ext}])$. As in the proof of Proposition \ref{prop:coin_wl}, let $\smash{[\bX^0, \tilde{\bX}^0]_{\text{swap}(S)}}$ be the matrix obtained from $[\bX^0, \tilde{\bX}^0]$ after swapping the column $\bX^0_j$ with the corresponding $\tilde{\bX}^0_j$ for all $j \in S$. Similarly, let $\smash{[\bX^{\ext}, \tilde{\bX}^{\ext}]_{\text{swap}(S)}}$ be the matrix obtained from $[\bX^{\ext}, \tilde{\bX}^{\ext}]$ after swapping the column $\bX^{\ext}_j$ with the corresponding $\tilde{\bX}^{\ext}_j$ for all $j \in S$. 
Consider swapping variables in $S$ in all environments. Let $\smash{W^{\wl}_{\text{swap}(S)} }= w^{\wl}(\bY^0, \smash{[\bX^0, \tilde{\bX}^0]_{\text{swap}(S)}}, \bY^{\ext}, \smash{[\bX^{\ext}, \tilde{\bX}^{\ext}]_{\text{swap}(S)}})$. 
On the one hand, since $\phi$ is a symmetric in $(\bX^0, \bX^{\ext})$ and $(\tilde{\bX}^0, \tilde{\bX}^{\ext})$, we have that
$\smash{W^{\wl}_{\text {swap}(S)}}=\varepsilon \odot W$. On the other hand, under the constant causal model, for any $j \in H_0$, $j$ is also null in other environments. Thus Lemma
1 in \citet{candes2018panning} implies that $[\bX^0, \tilde{\bX}^0] \, \smash{\stackrel{d}{=}} \, [\bX^0, \tilde{\bX}^0]_{\text{swap}(S)} \mid \bY^0$ and $[\bX^{\ext}, \tilde{\bX}^{\ext}] \, \smash{\stackrel{d}{=}} \, [\bX^{\ext}, \tilde{\bX}^{\ext}]_{\text{swap}(S)} \mid \bY^{\ext}$. 
Together with the fact that $([\bX^0, \tilde{\bX}^0], \bY^0)$ is independent of $([\bX^{\ext}, \tilde{\bX}^{\ext}], \bY^{\ext})$, the above results imply that 
\[(\bY^0, [\bX^0, \tilde{\bX}^0], \bY^{\ext}, [\bX^{\ext}, \tilde{\bX}^{\ext}])
\stackrel{d}{=}
(\bY^0, \smash{[\bX^0, \tilde{\bX}^0]_{\text{swap}(S)}}, \bY^{\ext}, \smash{[\bX^{\ext}, \tilde{\bX}^{\ext}]_{\text{swap}(S)}}).
\]
Thus, $W^{\wl}_{\text{swap}(S)} \, \smash{\stackrel{d}{=}} \, W^{\wl}$. These two results give the desired property of $\smash{W^{\wl} \stackrel{d}{=} \varepsilon \odot W^{\wl}}$. 

\FloatBarrier

\clearpage

\section{Additional results from the numerical experiments} \label{app:numerical-exp}

\begin{figure}[!htb]
  \centering
 \includegraphics[width=0.9\linewidth]{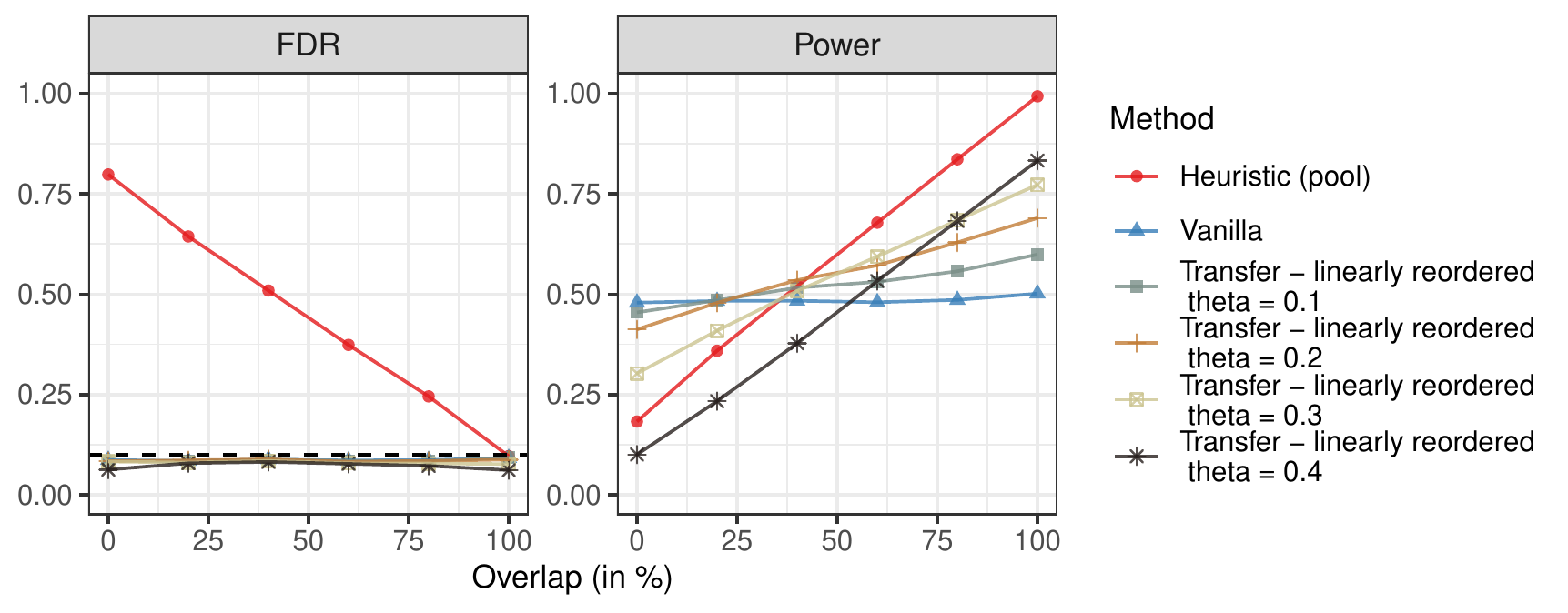}
 \caption{Performance of knockoffs with linear combination statistics, compared to benchmarks. We vary $\theta$ from $0.1$ to $0.4$. Each point averages the results of 500 independent experiments. Other details are as in Figure~\ref{fig:transfer_a}.}
 \label{fig:transfer_b}
\end{figure}

\begin{figure}[!htb]
 \centering
 \includegraphics[width=0.8\linewidth]{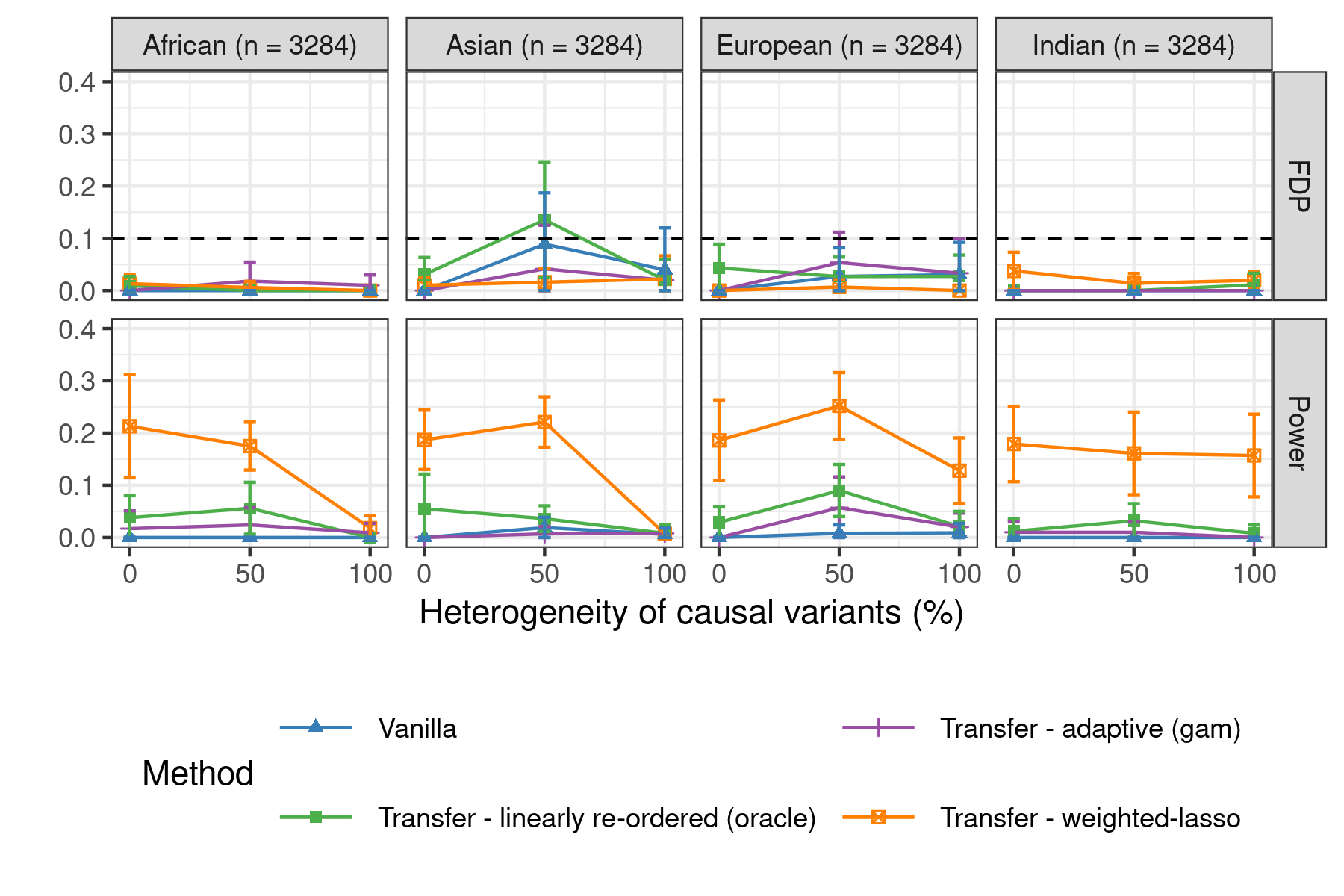}
 \caption{Comparison of the performances of different knockoff methods for transfer learning applied to simulated GWAS with real genotypes from different populations with equal sample sizes. Other details are as in Figure~\ref{fig:sim_ukb}. Note that the performance of the weighted-lasso transfer learning method is highest in the European population (and to a lesser extent also in the Indian population) when the heterogeneity of the causal variants is large, consistently with the fact that these groups are more closely related to the British one.}
 \label{fig:sim_ukb_small}
\end{figure}

\begin{figure}[!htb]
 \centering
 \includegraphics[width=\linewidth]{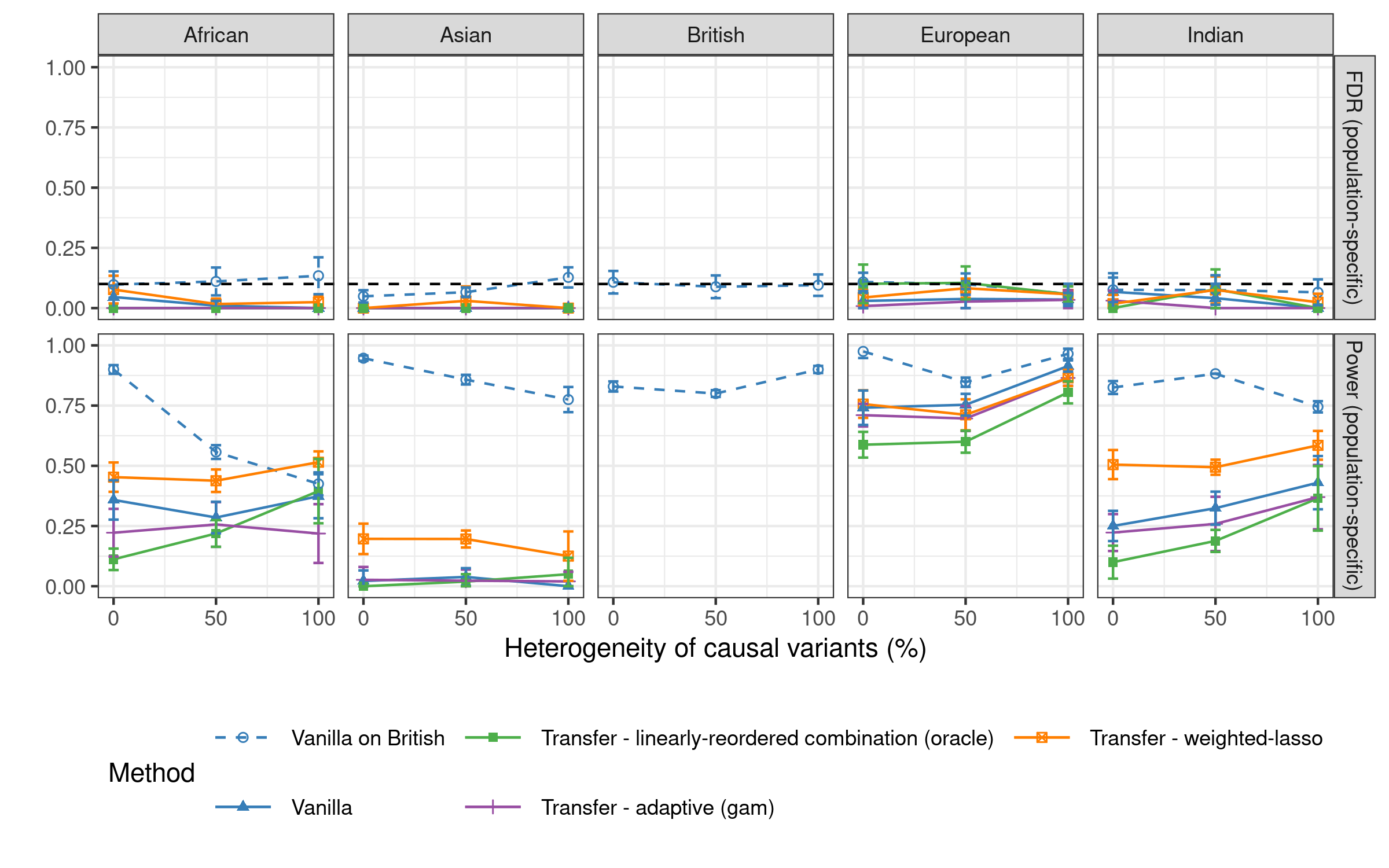}
 \caption{Comparison of the population-specific performances of different knockoff methods for transfer learning in the setting of Figure~\ref{fig:sim_ukb_specific}. The signal-to-noise ratio is 10\% here.}
 \label{fig:sim_ukb_specific_10}
\end{figure}

\begin{figure}[!htb]
 \centering
 \includegraphics[width=\linewidth]{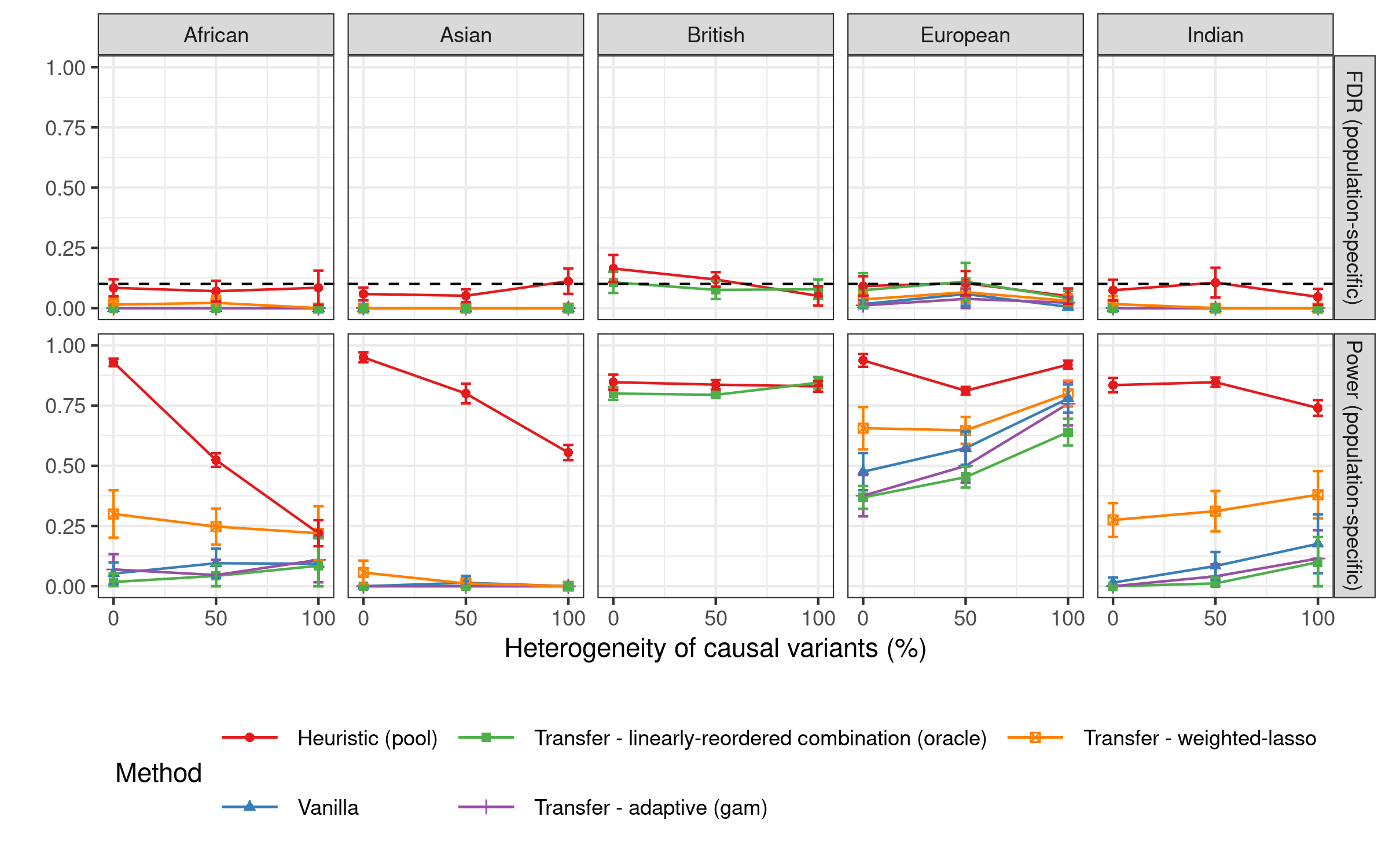}
 \caption{Additional comparison of the population-specific performances of different knockoff methods for transfer learning in the setting of Figure~\ref{fig:sim_ukb_specific}.}
 \label{fig:sim_ukb_specific_pooled}
\end{figure}

\FloatBarrier

\section{Additional results from the data analysis} \label{app:numerical-data}

\begin{table}[!htb]
\centering
\begin{tabular}{l|c|cccc}
\toprule
Population & Resolution (kb) & \makecell{Vanilla} 
           & \makecell{Transfer -\\ linearly re-ordered\\ ($\theta=0.1$)} 
           & \makecell{Transfer -\\ adaptive (gam)} 
           & \makecell{Transfer -\\ weighted-lasso}\\
\midrule
 & 3 & 15 & 13 & 20 & 37\\
%\cmidrule{2-6}
 & 20 & 26 & 51 & 46 & 52\\
%\cmidrule{2-6}
 & 41 & 75 & 77 & 57 & 64\\
%\cmidrule{2-6}
 & 81 & 94 & 113 & 123 & 204\\
%\cmidrule{2-6}
 & 208 & 113 & 154 & 216 & 350\\
%\cmidrule{2-6}
\multirow{-6}{*}{\raggedright\arraybackslash European} & 425 & 174 & 232 & 303 & 513\\
\cmidrule{1-6}
 & 3 & 5 & 3 & 3 & 1\\
%\cmidrule{2-6}
 & 20 & 2 & 4 & 10 & 8\\
%\cmidrule{2-6}
 & 41 & 2 & 4 & 6 & 11\\
%\cmidrule{2-6}
 & 81 & 6 & 14 & 17 & 8\\
%\cmidrule{2-6}
 & 208 & 3 & 12 & 17 & 13\\
%\cmidrule{2-6}
\multirow{-6}{*}{\raggedright\arraybackslash Indian} & 425 & 8 & 16 & 9 & 43\\
\cmidrule{1-6}
 & 3 & 4 & 5 & 5 & 3\\
%\cmidrule{2-6}
 & 20 & 2 & 1 & 0 & 22\\
%\cmidrule{2-6}
 & 41 & 1 & 1 & 1 & 15\\
%\cmidrule{2-6}
 & 81 & 0 & 0 & 0 & 6\\
%\cmidrule{2-6}
 & 208 & 7 & 15 & 13 & 25\\
%\cmidrule{2-6}
\multirow{-6}{*}{\raggedright\arraybackslash African} & 425 & 2 & 3 & 0 & 24\\
\cmidrule{1-6}
 & 3 & 3 & 3 & 1 & 1\\
%\cmidrule{2-6}
 & 20 & 3 & 0 & 0 & 11\\
%\cmidrule{2-6}
 & 41 & 2 & 4 & 5 & 4\\
%\cmidrule{2-6}
 & 81 & 3 & 3 & 5 & 16\\
%\cmidrule{2-6}
 & 208 & 3 & 3 & 1 & 7\\
%\cmidrule{2-6}
\multirow{-6}{*}{\raggedright\arraybackslash Asian} & 425 & 0 & 0 & 0 & 0\\
\bottomrule
\end{tabular}
\caption{Numbers of discoveries for height reported at different resolution by vanilla knockoffs and three transfer learning methods applied to data from different minority populations in the UK Biobank. Other details are as in Table~\ref{tab:platelet}.}
\label{tab:height}
\end{table}

\begin{table}[!htb]
\centering
\begin{tabular}{l|c|cccc}
\toprule
Population & Resolution (kb) & \makecell{Vanilla} 
           & \makecell{Transfer -\\ linearly re-ordered\\ ($\theta = 0.1$)}
           & \makecell{Transfer -\\ adaptive (gam)} 
           & \makecell{Transfer -\\ weighted-lasso} \\
\midrule
 & 3 & 1 & 1 & 1 & 0\\
%\cmidrule{2-6}
 & 20 & 6 & 8 & 10 & 7\\
%\cmidrule{2-6}
 & 41 & 7 & 7 & 1 & 11\\
%\cmidrule{2-6}
 & 81 & 3 & 3 & 13 & 21\\
%\cmidrule{2-6}
 & 208 & 9 & 12 & 7 & 25\\
%\cmidrule{2-6}
\multirow{-6}{*}{\raggedright\arraybackslash European} & 425 & 7 & 10 & 19 & 47\\
\cmidrule{1-6}
 & 3 & 7 & 8 & 8 & 5\\
%\cmidrule{2-6}
 & 20 & 7 & 10 & 10 & 5\\
%\cmidrule{2-6}
 & 41 & 3 & 3 & 10 & 5\\
%\cmidrule{2-6}
 & 81 & 7 & 9 & 10 & 14\\
%\cmidrule{2-6}
 & 208 & 6 & 11 & 4 & 13\\
%\cmidrule{2-6}
\multirow{-6}{*}{\raggedright\arraybackslash Indian} & 425 & 4 & 4 & 6 & 7\\
\cmidrule{1-6}
 & 3 & 0 & 0 & 0 & 1\\
%\cmidrule{2-6}
 & 20 & 0 & 0 & 0 & 0\\
%\cmidrule{2-6}
 & 41 & 1 & 1 & 0 & 4\\
%\cmidrule{2-6}
 & 81 & 1 & 1 & 2 & 4\\
%\cmidrule{2-6}
 & 208 & 0 & 0 & 0 & 3\\
%\cmidrule{2-6}
\multirow{-6}{*}{\raggedright\arraybackslash African} & 425 & 1 & 1 & 0 & 8\\
\cmidrule{1-6}
 & 3 & 2 & 2 & 2 & 0\\
%\cmidrule{2-6}
 & 20 & 1 & 1 & 1 & 8\\
%\cmidrule{2-6}
 & 41 & 2 & 2 & 2 & 1\\
%\cmidrule{2-6}
 & 81 & 1 & 1 & 2 & 7\\
%\cmidrule{2-6}
 & 208 & 0 & 0 & 0 & 0\\
%\cmidrule{2-6}
\multirow{-6}{*}{\raggedright\arraybackslash Asian} & 425 & 1 & 1 & 2 & 2\\
\bottomrule
\end{tabular}
\caption{Numbers of discoveries for BMI reported at different resolution by vanilla knockoffs and three transfer learning methods applied to data from different minority populations in the UK Biobank. Other details are as in Table~\ref{tab:platelet}.}
\label{tab:bmi}
\end{table}

\begin{table}[!htb]
\begin{tabular}[t]{l|c|cccc}
\toprule
Population & Resolution & Vanilla 
           & \makecell{Transfer -\\ adaptive (gam)} 
           & \makecell{Transfer -\\ weighted-lasso}
           & \makecell{Transfer -\\ multi-weighted-lasso}\\
\midrule
 & 3 & 1 & 1 & 0 & 0\\
%\cmidrule{2-6}
 & 20 & 6 & 7 & 7 & 8\\
%\cmidrule{2-6}
 & 41 & 7 & 13 & 11 & 10\\
%\cmidrule{2-6}
 & 81 & 3 & 2 & 21 & 2\\
%\cmidrule{2-6}
 & 208 & 9 & 16 & 25 & 12\\
%\cmidrule{2-6}
\multirow{-6}{*}{\raggedright\arraybackslash European} & 425 & 7 & 50 & 47 & 22\\
\cmidrule{1-6}
 & 3 & 7 & 0 & 5 & 2\\
%\cmidrule{2-6}
 & 20 & 7 & 5 & 5 & 4\\
%\cmidrule{2-6}
 & 41 & 3 & 5 & 5 & 4\\
%\cmidrule{2-6}
 & 81 & 7 & 13 & 14 & 7\\
%\cmidrule{2-6}
 & 208 & 6 & 7 & 13 & 9\\
%\cmidrule{2-6}
\multirow{-6}{*}{\raggedright\arraybackslash Indian} & 425 & 4 & 1 & 7 & 5\\
\cmidrule{1-6}
 & 3 & 0 & 2 & 1 & 0\\
%\cmidrule{2-6}
 & 20 & 0 & 0 & 0 & 1\\
%\cmidrule{2-6}
 & 41 & 1 & 0 & 4 & 0\\
%\cmidrule{2-6}
 & 81 & 1 & 0 & 4 & 4\\
%\cmidrule{2-6}
 & 208 & 0 & 0 & 3 & 0\\
%\cmidrule{2-6}
\multirow{-6}{*}{\raggedright\arraybackslash African} & 425 & 1 & 0 & 8 & 0\\
\cmidrule{1-6}
 & 3 & 2 & 0 & 0 & 2\\
%\cmidrule{2-6}
 & 20 & 1 & 1 & 8 & 3\\
%\cmidrule{2-6}
 & 41 & 2 & 3 & 1 & 2\\
%\cmidrule{2-6}
 & 81 & 1 & 1 & 7 & 1\\
%\cmidrule{2-6}
 & 208 & 0 & 0 & 0 & 1\\
%\cmidrule{2-6}
\multirow{-6}{*}{\raggedright\arraybackslash Asian} & 425 & 1 & 3 & 2 & 2\\
\bottomrule
\end{tabular}
\caption{Numbers of discoveries for BMI, using UK Biobank data from different minority populations and prior knowledge about four related phenotypes acquired from the analysis of British individuals. Other details are as in Table~\ref{tab:multi_bmi_small}.
}
\label{tab:multi_bmi}
\end{table}

\begin{table}[!htb]
\begin{tabular}[t]{l|c|cccc}
\toprule
Population & Resolution & Vanilla 
           & \makecell{Transfer -\\ adaptive (gam)}
           & \makecell{Transfer -\\ weighted-lasso} 
           & \makecell{Transfer -\\ multi-weighted-lasso}\\
\midrule
 & 3 & 0 & 1 & 1 & 0\\
%\cmidrule{2-6}
 & 20 & 4 & 10 & 8 & 12\\
%\cmidrule{2-6}
 & 41 & 1 & 2 & 3 & 2\\
%\cmidrule{2-6}
 & 81 & 2 & 3 & 3 & 10\\
%\cmidrule{2-6}
 & 208 & 1 & 3 & 3 & 1\\
%\cmidrule{2-6}
\multirow{-6}{*}{\raggedright\arraybackslash European} & 425 & 2 & 14 & 5 & 4\\
\cmidrule{1-6}
 & 3 & 1 & 3 & 2 & 1\\
%\cmidrule{2-6}
 & 20 & 1 & 2 & 3 & 5\\
%\cmidrule{2-6}
 & 41 & 1 & 0 & 3 & 3\\
%\cmidrule{2-6}
 & 81 & 1 & 4 & 2 & 1\\
%\cmidrule{2-6}
 & 208 & 0 & 0 & 2 & 1\\
%\cmidrule{2-6}
\multirow{-6}{*}{\raggedright\arraybackslash Indian} & 425 & 1 & 0 & 3 & 3\\
\cmidrule{1-6}
 & 3 & 1 & 0 & 3 & 2\\
%\cmidrule{2-6}
 & 20 & 1 & 1 & 2 & 2\\
%\cmidrule{2-6}
 & 41 & 0 & 2 & 2 & 1\\
%\cmidrule{2-6}
 & 81 & 1 & 1 & 2 & 2\\
%\cmidrule{2-6}
 & 208 & 1 & 0 & 2 & 2\\
%\cmidrule{2-6}
\multirow{-6}{*}{\raggedright\arraybackslash African} & 425 & 1 & 5 & 2 & 2\\
\cmidrule{1-6}
 & 3 & 1 & 0 & 2 & 0\\
%\cmidrule{2-6}
 & 20 & 0 & 2 & 0 & 0\\
%\cmidrule{2-6}
 & 41 & 0 & 0 & 0 & 0\\
%\cmidrule{2-6}
 & 81 & 0 & 0 & 0 & 0\\
%\cmidrule{2-6}
 & 208 & 0 & 0 & 0 & 2\\
%\cmidrule{2-6}
\multirow{-6}{*}{\raggedright\arraybackslash Asian} & 425 & 1 & 1 & 1 & 2\\
\bottomrule
\end{tabular}
\caption{Numbers of discoveries for diabetes, using UK Biobank data from different minority populations and prior knowledge about four related phenotypes acquired from the analysis of British individuals. Other details are as in Table~\ref{tab:multi_bmi_small}.}
\label{tab:multi_diabetes}
\end{table}

\begin{table}[!htb]
\begin{tabular}[t]{l|c|cccc}
\toprule
Population & Resolution & Vanilla 
           & \makecell{Transfer -\\ adaptive (gam)}
           & \makecell{Transfer -\\ weighted-lasso}
           & \makecell{Transfer -\\ multi-weighted-lasso}\\
\midrule
 & 3 & 1 & 3 & 3 & 4\\
%\cmidrule{2-6}
 & 20 & 3 & 5 & 5 & 5\\
%\cmidrule{2-6}
 & 41 & 2 & 9 & 8 & 6\\
%\cmidrule{2-6}
 & 81 & 2 & 21 & 20 & 15\\
%\cmidrule{2-6}
 & 208 & 2 & 23 & 32 & 10\\
%\cmidrule{2-6}
\multirow{-6}{*}{\raggedright\arraybackslash European} & 425 & 5 & 31 & 31 & 22\\
\cmidrule{1-6}
 & 3 & 0 & 1 & 0 & 0\\
%\cmidrule{2-6}
 & 20 & 2 & 0 & 0 & 0\\
%\cmidrule{2-6}
 & 41 & 1 & 1 & 0 & 0\\
%\cmidrule{2-6}
 & 81 & 1 & 3 & 2 & 5\\
%\cmidrule{2-6}
 & 208 & 7 & 0 & 0 & 3\\
%\cmidrule{2-6}
\multirow{-6}{*}{\raggedright\arraybackslash Indian} & 425 & 3 & 0 & 0 & 1\\
\cmidrule{1-6}
 & 3 & 1 & 4 & 1 & 1\\
%\cmidrule{2-6}
 & 20 & 1 & 1 & 1 & 0\\
%\cmidrule{2-6}
 & 41 & 1 & 1 & 0 & 1\\
%\cmidrule{2-6}
 & 81 & 1 & 2 & 2 & 2\\
%\cmidrule{2-6}
 & 208 & 1 & 0 & 1 & 2\\
%\cmidrule{2-6}
\multirow{-6}{*}{\raggedright\arraybackslash African} & 425 & 1 & 0 & 1 & 1\\
\cmidrule{1-6}
 & 3 & 0 & 4 & 0 & 0\\
%\cmidrule{2-6}
 & 20 & 0 & 0 & 0 & 0\\
%\cmidrule{2-6}
 & 41 & 0 & 1 & 0 & 0\\
%\cmidrule{2-6}
 & 81 & 0 & 0 & 0 & 0\\
%\cmidrule{2-6}
 & 208 & 4 & 0 & 0 & 0\\
%\cmidrule{2-6}
\multirow{-6}{*}{\raggedright\arraybackslash Asian} & 425 & 0 & 0 & 0 & 0\\
\bottomrule
\end{tabular}
\caption{Numbers of discoveries for systolic blood pressure, using UK Biobank data from different minority populations and prior knowledge about four related phenotypes acquired from the analysis of British individuals. Other details are as in Table~\ref{tab:multi_bmi_small}.}
\label{tab:multi_sbp}
\end{table}

\begin{table}[!htb]
\begin{tabular}[t]{l|c|cccc}
\toprule
Population & Resolution & Vanilla 
           & \makecell{Transfer -\\ adaptive (gam)}
           & \makecell{Transfer -\\ weighted-lasso}
           & \makecell{Transfer -\\ multi-weighted-lasso}\\
\midrule
 & 3 & 4 & 1 & 1 & 4\\
%\cmidrule{2-6}
 & 20 & 0 & 2 & 3 & 1\\
%\cmidrule{2-6}
 & 41 & 0 & 3 & 3 & 2\\
%\cmidrule{2-6}
 & 81 & 4 & 6 & 7 & 3\\
%\cmidrule{2-6}
 & 208 & 1 & 2 & 2 & 10\\
%\cmidrule{2-6}
\multirow{-6}{*}{\raggedright\arraybackslash European} & 425 & 1 & 4 & 4 & 2\\
\cmidrule{1-6}
 & 3 & 0 & 1 & 0 & 0\\
%\cmidrule{2-6}
 & 20 & 0 & 0 & 0 & 0\\
%\cmidrule{2-6}
 & 41 & 0 & 0 & 0 & 0\\
%\cmidrule{2-6}
 & 81 & 0 & 0 & 0 & 0\\
%\cmidrule{2-6}
 & 208 & 0 & 2 & 1 & 2\\
%\cmidrule{2-6}
\multirow{-6}{*}{\raggedright\arraybackslash Indian} & 425 & 1 & 2 & 2 & 3\\
\cmidrule{1-6}
 & 3 & 1 & 0 & 2 & 3\\
%\cmidrule{2-6}
 & 20 & 2 & 2 & 1 & 2\\
%\cmidrule{2-6}
 & 41 & 2 & 0 & 1 & 1\\
%\cmidrule{2-6}
 & 81 & 0 & 0 & 0 & 0\\
%\cmidrule{2-6}
 & 208 & 0 & 7 & 0 & 0\\
%\cmidrule{2-6}
\multirow{-6}{*}{\raggedright\arraybackslash African} & 425 & 0 & 4 & 0 & 0\\
\cmidrule{1-6}
 & 3 & 1 & 3 & 4 & 5\\
%\cmidrule{2-6}
 & 20 & 1 & 0 & 2 & 0\\
%\cmidrule{2-6}
 & 41 & 0 & 0 & 0 & 0\\
%\cmidrule{2-6}
 & 81 & 0 & 0 & 0 & 0\\
%\cmidrule{2-6}
 & 208 & 0 & 1 & 0 & 0\\
%\cmidrule{2-6}
\multirow{-6}{*}{\raggedright\arraybackslash Asian} & 425 & 4 & 0 & 3 & 1\\
\bottomrule
\end{tabular}
\caption{Numbers of discoveries for cardiovascular disease, using UK Biobank data from different minority populations and prior knowledge about four related phenotypes acquired from the analysis of British individuals. Other details are as in Table~\ref{tab:multi_bmi_small}.}
\label{tab:multi_cvd}
\end{table}

\end{document}